\documentclass[conference]{IEEEtran}

\usepackage{algorithm}
\usepackage{algorithmicx}
\usepackage[noend]{algpseudocode}
\algnewcommand\algorithmicforeach{\textbf{for each}}
\algdef{S}[FOR]{ForEach}[1]{\algorithmicforeach\ #1\ \algorithmicdo}

\usepackage{textcomp}
\usepackage{svg}
\usepackage{amsmath}
\usepackage{xcolor}
\usepackage{graphicx}
\usepackage{graphics}
\usepackage{float}
\usepackage{cclicenses}
\usepackage{xspace}
\usepackage{subfigure}
\usepackage{makecell}
\usepackage{multirow} 

\begin{document}

\title{Demystifying the Performance of Data Transfers in High-Performance Research Networks}


\author{
    \IEEEauthorblockN{Ehsan Saeedizade}
    \IEEEauthorblockA{University of Nevada, Reno\\
    ehsansaeedizade@nevada.unr.edu} 
    
    \and \IEEEauthorblockN{Bing Zhang}
    \IEEEauthorblockA{National Center for Supercomputing Applications\\ bing@illinois.edu}
    
    \and \IEEEauthorblockN{Engin Arslan}
    \IEEEauthorblockA{University of Texas at Arlington\\ engin.arslan@uta.edu}
}

\maketitle 

\begin{abstract}
High-speed research networks are built to meet the ever-increasing needs of data-intensive distributed workflows. However, data transfers in these networks often fail to attain the promised transfer rates for several reasons, including I/O and network interference, server misconfigurations, and network anomalies. Although understanding the root causes of performance issues is critical to mitigating them and increasing the utilization of expensive network infrastructures, there is currently no available mechanism to monitor data transfers in these networks. In this paper, we present a scalable, end-to-end monitoring framework to gather and store key performance metrics for file transfers to shed light on the performance of transfers. The evaluation results show that the proposed framework can monitor up to $400$ transfers per host and more than $40,000$ transfers in total while collecting performance statistics at one-second precision. We also introduce a heuristic method to automatically process the gathered performance metrics and identify the root causes of performance anomalies with an F-score of $87-98\%$. 
\end{abstract}

\maketitle

\section{Introduction}
Data volumes generated by scientific applications are increasing faster than ever. For instance, the cosmology and astronomical project Square Kilometer Array produces one exabyte of data every $13$ days~\cite{kettimuthu2014modeling}. This massive volume of data often needs to be transferred to remote sites for different reasons, such as processing and long-term archival. Significant investments have been made to build high-performance wide-area research networks (e.g., Internet-2 and ESnet) with up to $400$Gbps bandwidth to facilitate these data movements. Yet, data transfers in these research networks often experience significant performance fluctuations due to various reasons, including I/O and network congestion, host misconfiguration, and hardware malfunctions.

\begin{figure}
\centering
    \includegraphics[width=0.8\linewidth]{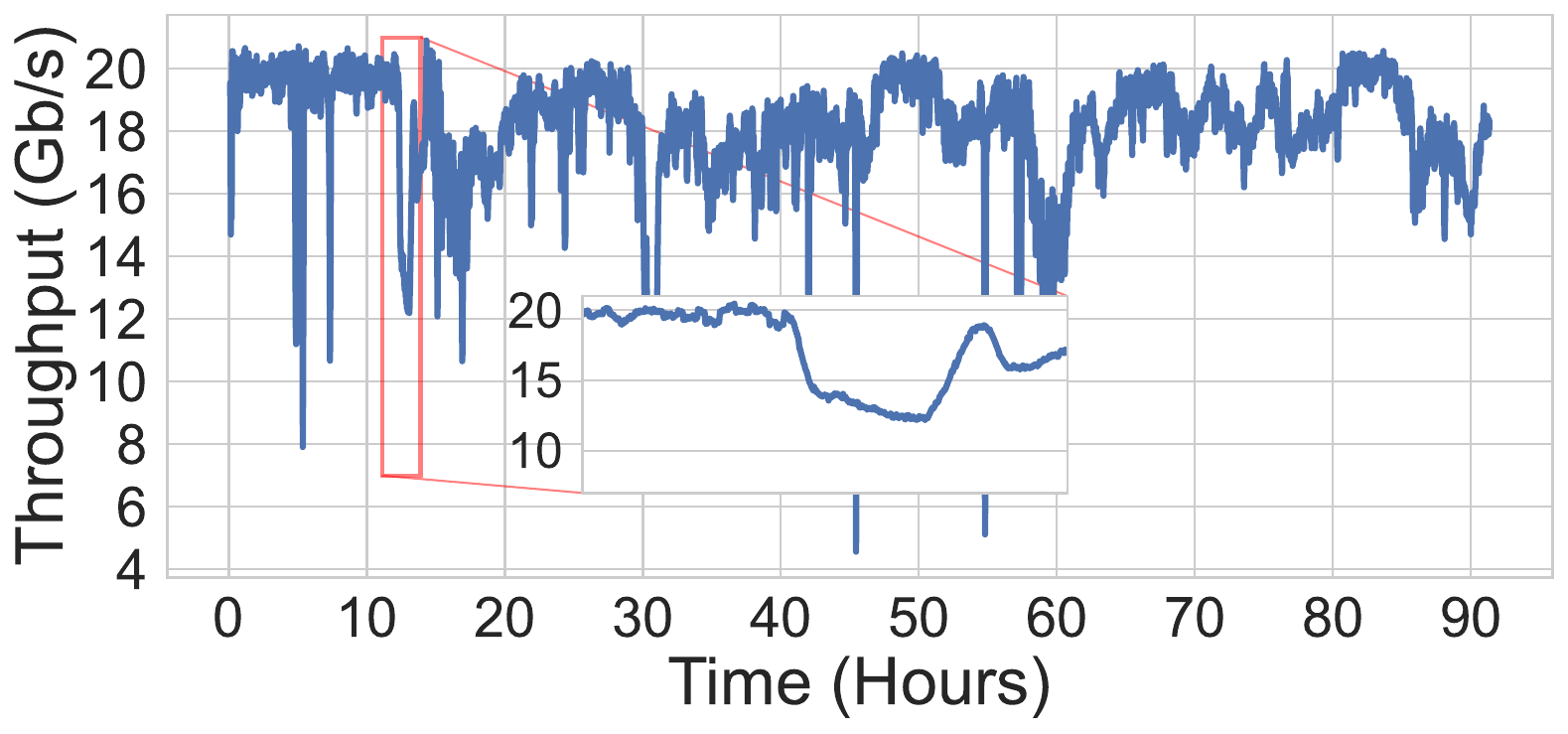} 
    \vspace{-3mm}
    \caption{Throughput of a long-running file transfer between Stampede2 and Expanse cluster exhibits significant fluctuations.}
    \label{fig:motivation}
    \vspace{-6mm}
\end{figure}

Figure~\ref{fig:motivation} shows the throughput of a long-running transfer job (consists of $1000\times1GiB$ files) between Stampede2~\cite{stampede2} and Expanse~\cite{expanse} supercomputers that are connected with $100$ Gbps network and $30$ms round trip time. Although both clusters have high-performance parallel file systems and dedicated data transfer nodes with more than $ 100$Gbps capacity, the transfer cannot attain more than $20$ Gbps throughput. Even worse, the throughput fluctuates significantly and falls below $5$ Gbps several times. While resource interference is the usual suspect, non-congestion-related issues such as hardware and software anomalies happen at a significant frequency. Identifying the root cause of performance issues (e.g., congestion at the source file system and insufficient TCP buffer size) is critical to the necessary actions to address the underlying reasons.

For system monitoring, HPC operators typically rely on general-purpose (e.g., Ganglia~\cite{ganglia}, Collectl~\cite{collectl}, IPM~\cite{fuerlinger2010effective}) or custom (e.g., OVIS\cite{brandt2013high}, ISC~\cite{semeraro2014takes}, and TACC Stats~\cite{hammond2011tacc_stats}) tools to collect performance counters of compute nodes, storage servers, and network interconnect. While these tools offer cluster-wide and application-level resource monitoring for compute jobs, they do not offer similar coverage for wide-area transfers that span multiple administrative domains. In many cases, even cluster-level resource usage of transfers (e.g., data transfer nodes and border routers) are not monitored by the cluster-wide system monitoring tools as these resources are deemed to be part of networking infrastructure and thus should be monitored by networking teams. Beyond the cluster boundaries, no solution exists to monitor the entire lifecycle of file data transfers in research networks, making it nearly impossible to reason about the performance issues. 
 
 Many research and education institutions use PerfSonar~\cite{hanemann2005perfsonar} to conduct transfers between endpoints periodically and identify performance anomalies. However, the active probing approach (usually once in $6-12$ hours) cannot detect anomalies between probing periods. Moreover, most institutions use PerfSonar to schedule memory-to-memory transfers, which cannot detect issues caused by file systems, such as I/O congestion. Similarly, port- (e.g., SNMP) and flow- (e.g., NetFlow) level metrics captured from network devices can help localize network issues such as malfunctioning ports and routing instabilities, but they cannot be used to explain end host and I/O-related issues. Moreover, port and flow level statistics may not be accessible since they are either not collected at high precision (e.g., typical SNMP data collection frequency ranges between $30$ seconds to $5$ minutes) or not shared due to privacy concerns. 

Providing insights into data transfer bottlenecks has several benefits to users and system administrators. \textit{Benefits to Users:} Scientific datasets are often accessible from multiple locations. For example, the Sequence Read Archive dataset can be downloaded from a central repository (e.g., NCBI) or cloud providers. Thus, if users can learn what causes performance bottlenecks when downloading datasets from a remote cluster, they can choose an endpoint with more capacity and less congestion. Moreover, users can utilize concurrent file transfers~\cite{arifuzzaman2021online,arifuzzaman2023falcon,arifuzzaman2023use} and congestion-aware I/O techniques to overcome I/O and network interference~\cite{kim2015lads}. \textit{Benefits to System Administrators:} As HPC facilities and networks are expensive investments, increasing their utilization is of utmost importance for investors. Hence, real-time monitoring of wide-area transfers will help system administrators quickly troubleshoot any performance problems, thereby improving the quality of experience for users. Moreover, it will ease the burden on system administrators as they do not have to go through multiple system logs (e.g., I/O and network logs) and contact multiple groups (e.g., storage and networking) to pinpoint the root causes of issues.

Therefore, this paper proposes a scalable, high-precision monitoring framework for wide-area file transfers. It utilizes lightweight monitoring agents running on data transfer nodes to detect new transfers and capture performance metrics from both ends of transfers. Captured statistics are then sent to a cloud-hosted data collector in real-time for real-time analysis and visualization. The monitoring agents adopt several optimizations, such as data prefetching and caching, to quickly capture performance metrics and lower the overhead of monitoring agents. We show that the proposed monitoring framework can monitor more than $40,000$ transfers spanned across many data transfer nodes. We further explored automated processing of captured metrics for root cause analysis using machine learning and heuristic models. We reproduced the eight most common transfer anomalies with varying severity levels in eight network settings. We observe that machine learning models perform well when the training data is available for every network. However, trained models perform poorly when they are transferred to new settings. Hence, we propose a heuristic method that can determine the root causes of the eight performance anomalies with over $85\%$ F-score in all networks. 
In summary, the main contributions of this paper are as follows:

\begin{enumerate}
    \item We identify key performance metrics for wide-area file transfers that can be collected at the end hosts to localize performance problems.
    \item We design and develop a highly scalable real-time monitoring framework to collect performance statistics for file transfer in real-time with high precision. We show that the proposed architecture can monitor up to $400$ transfers running in the same node and $40,000$ transfers in total.
    \item We reproduce eight common performance anomalies in eight different networks to train ML and heuristic models to identify the root causes of transfer issues. We find that the heuristic model attains $87-98\%$ F-score in all eight networks. 
\end{enumerate}


The rest of this paper is organized as follows:  We discuss related work in Section II and background and motivation in Section III. The details of the proposed solution are presented in Section IV. We present experimental results in Section V and summarize the findings in Section VI.

\section{Related Work}
Researchers proposed several solutions to detect anomalies in data transfers, including supervised~\cite{kandula2009detailed, bahl2007towards} and unsupervised ~\cite{lakhina2004diagnosing,rao2018detecting, logg2004experiences} machine learning and heuristic models~\cite{arifuzzaman2021learning,arifuzzaman2021towards}. These solutions, however, only aimed at detecting or pinpointing the root causes of network-related issues, thus, they are not suited for file transfers that involve I/O operations. Traverso et al.~\cite{traverso2014exploiting} used active and passive measurements together to collect several transfer metrics, which are then analyzed manually to detect abnormal transfer behavior, such as high packet loss and retransmission rates. 
Mellia et al.~\cite{mellia2006passive} proposed a heuristic classification technique for anomalies that may occur in TCP connections using out-of-sequence and duplicate packets. 

Lakhina et al. used principal component analysis on link utilization data (i.e., SNMP) to detect volume-related anomalies~\cite{lakhina2004diagnosing}. They proposed a threshold-based detection algorithm to identify significant variance in {\em residual} traffic and distinguish the flows involved in the anomalous event.
Similarly, Rao et al. used principal component analysis on TCP performance metrics to extract principal components for normal traffic in high-speed networks~\cite{rao2018detecting}. Mapping normal and anomalous traffic into a new space using the first two principal components exposes the difference between these normal and anomalous groups since features found for normal traffic have different distributions for abnormal traffic. In another recent work, Ana et al. proposed FlowZilla, which detects transfer throughput anomalies by analyzing the flow size~\cite{anna2018flowzilla}. FlowZilla captures flow statistics using the host-based network monitoring tool {\em Tstat}~\cite{mellia2003tstat} and uses a Random Forest Regressor to relate them to the flow size. Once the flow size of traffic is estimated, it uses an adaptive threshold approach to determine if the volume of a flow is too small or large, which is then used to infer anomalies. 



\begin{figure*}
\begin{center}
    \hspace{-3mm}
    \subfigure[Network Anomalies]{
    \includegraphics[keepaspectratio=true,angle=0,width=.32\textwidth]{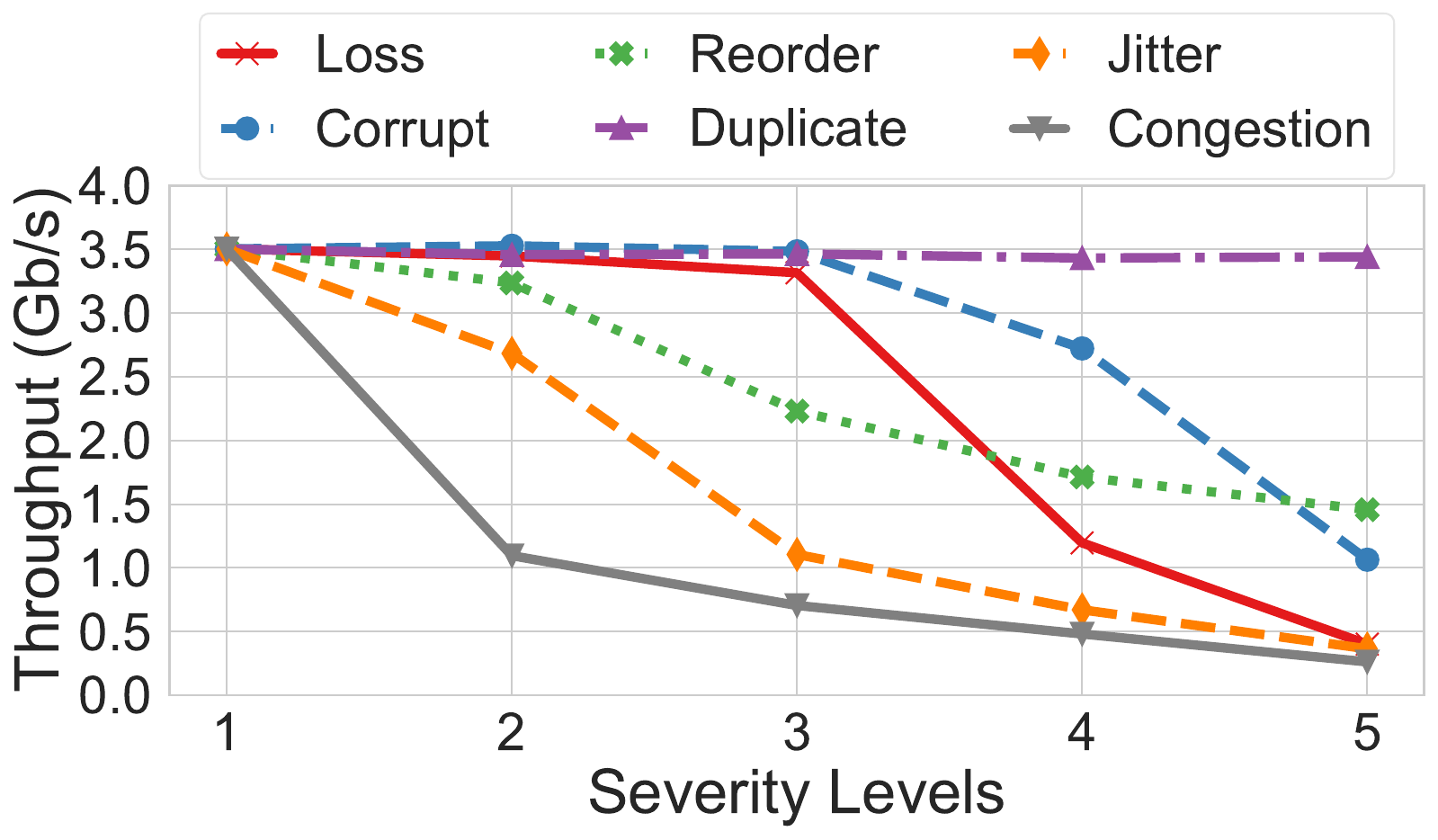}
    \label{fig:net_cong_a}}
    \hspace{-3mm}
    \subfigure[I/O Congestion]{
    \includegraphics[keepaspectratio=true,angle=0,width=.32\textwidth] {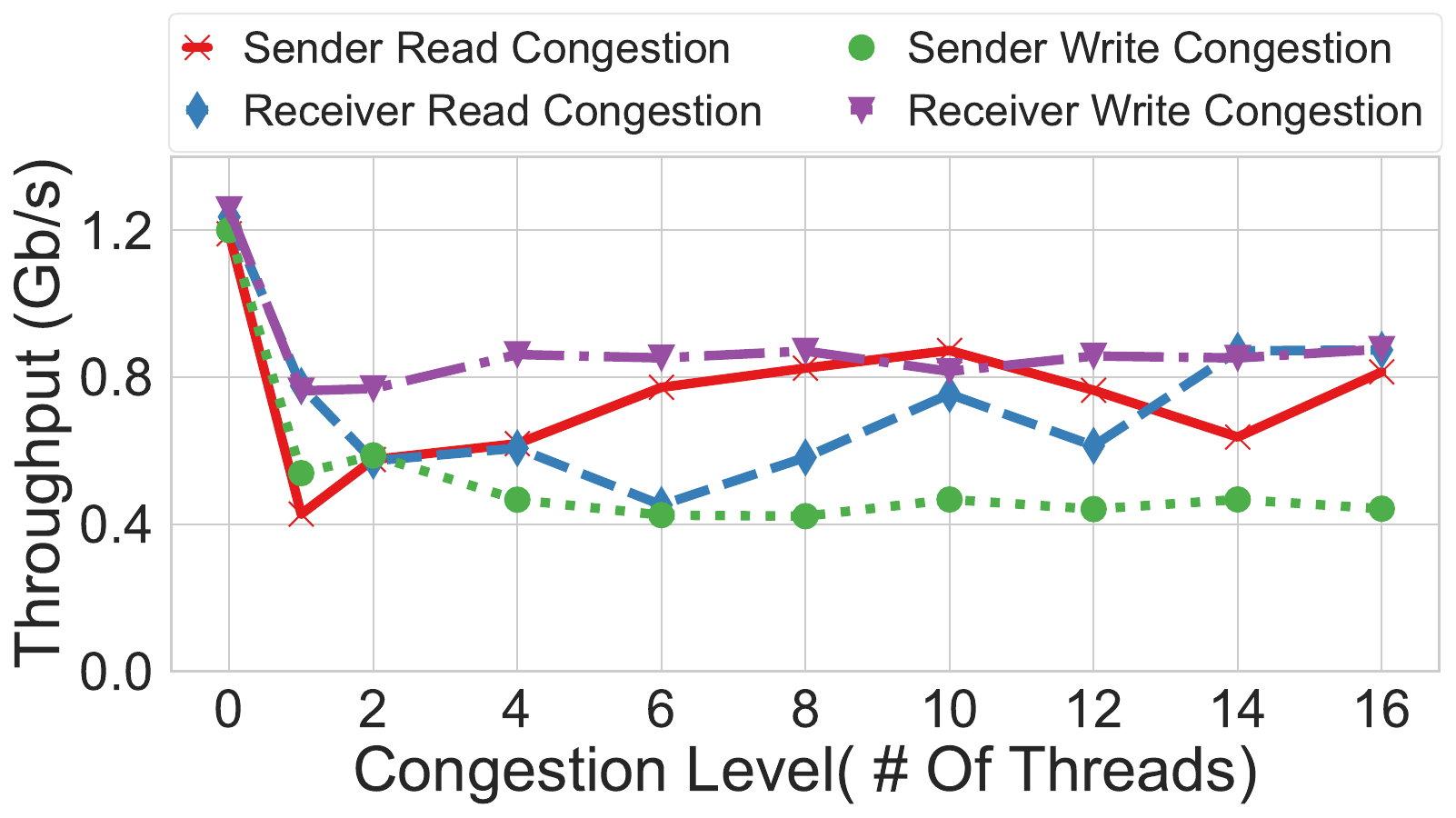}
    \label{fig:read_write_cong}}
    \hspace{-3mm}
    \subfigure[TCP Buffer Size Limitation]{
    \includegraphics[keepaspectratio=true,angle=0,width=.32\textwidth] {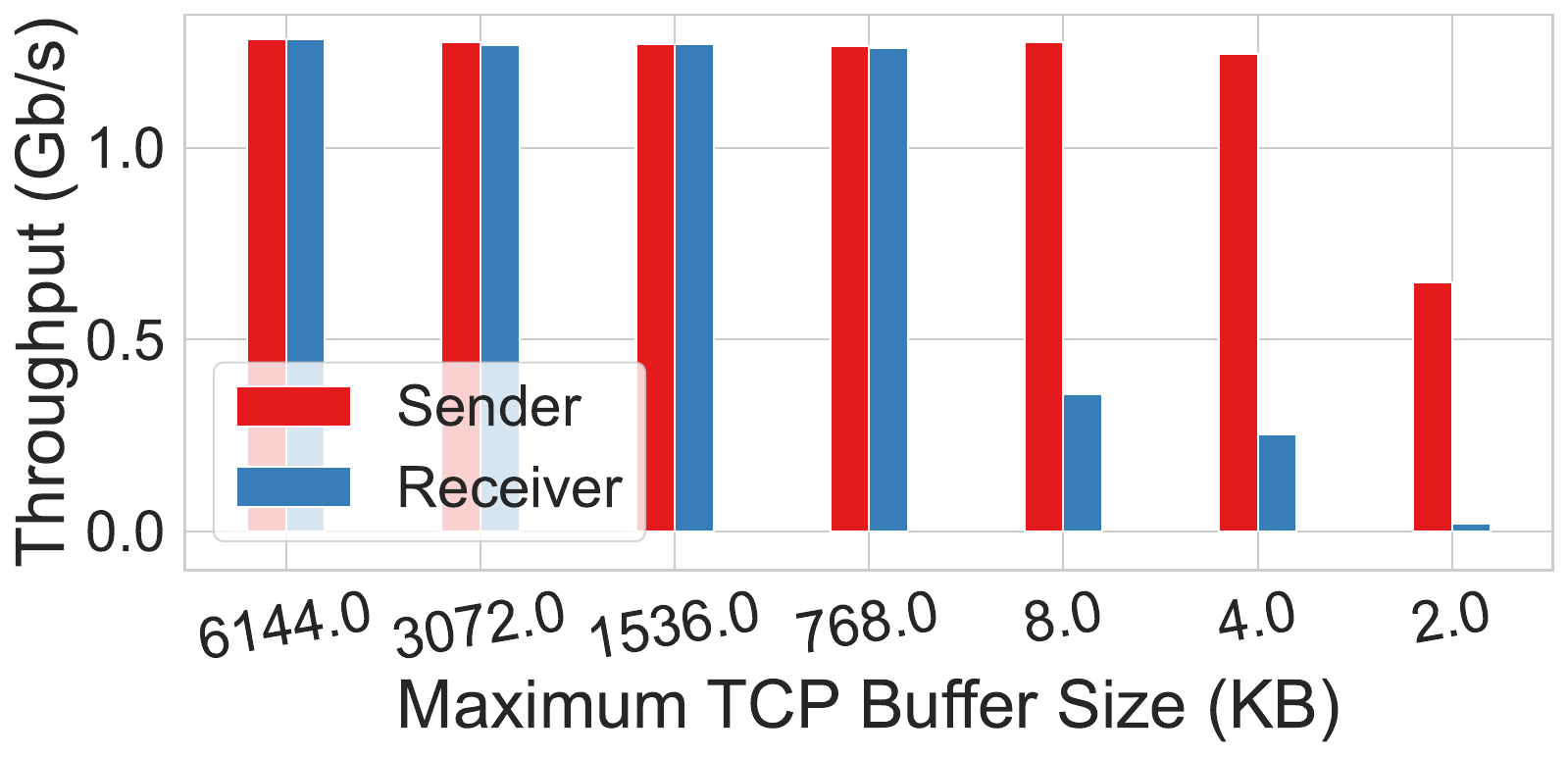}
    \label{fig:tcp_buff}}
\caption{File transfer throughput is affected by network congestion (a), I/O interference (b), and DTN misconfigurations (c); thus, it is key to gather performance statistics from all involved components to identify the root causes of performance issues. }
\vspace{-5mm}
\label{fig:anomaly_examples}
\end{center}
\end{figure*} 

Maassen et al.~\cite{maassen2006middleware} proposed a unified interface, Delphoi, to monitor grid resources. Delphoi only captures network statistics for transfers; thus, it is not designed to monitor file transfers.  Andreeva et al. developed a tool to monitor heterogeneous data transfers that use different technologies~\cite{andreeva2014wlcg}. Their work differs from ours because they only report general information, such as the throughput and the technology of the transfers. In this work, we proposed a framework that monitors the end-to-end performance of data transfers and collects and reports low-level metrics from all components involved in a transfer. Similarly, Aimar et al. proposed a Unified Monitoring Architecture to collect, store, search, and process monitoring data from various sources. However, their work does not directly collect performance metrics for data transfer or try to predict the root cause of the performance degradation of the transfers~\cite{aimar2017unified}.
Finally, Liu et al. analyzed historical Globus transfer logs to retrospectively predict the throughput of transfers based on transfer settings and the presence of other Globus transfers~\cite{liu2017explaining}. In other words, they estimate the throughput of completed Globus transfers, whereas this work presents a real-time monitoring framework for file transfers such that performance anomalies can be detected quickly.

\section{Background and Motivation}

File transfers involve three main steps as read operation on the source site, network transfer over the wide area network, and write operation on the destination side. Hence, transfer throughput is limited by the slowest of these three operations. A previous study showed by analyzing $30K$ Globus transfers that resource interference (both at the file system and network level) is the main source of performance limitations for file transfers in high-performance networks~\cite{liu2017explaining}. Hence, we identify five main components that resource interference can take place as source file system, source DTN, network, destination DTN, and destination file system. Thus, we first present common bottleneck scenarios that can happen in each of these components. 

\textbf{Wide Area Network:} Even if HPC clusters are connected with dedicated high-speed networks (e.g., Internet-2 and ESnet), network anomalies and congestion can limit the throughput of transfers in these networks. As an example, a faulty network interface card can drop or corrupt network packets, causing congestion control algorithms to lower sending rate as commonly used congestion control protocols relate packet losses to network congestion. Thus, we emulated five different network issues as packet loss, packet corruption, packet reordering, packet duplication, and delay jitter using Linux's \texttt{netem} utility. \emph{Loss} anomaly drops packets randomly while \emph{corrupt} anomaly alters packet content to cause TCP checksum mismatch. Loss and corrupt anomalies can take place due to faulty network devices and firmware bugs. \emph{Reorder} anomaly changes the order of some packets to emulate a scenario in which different packets take different routes while being transmitted, thus they arrive in a different order than they are sent. \emph{Duplicate} anomaly causes some packets to be sent to the receiver multiple times either by the sender or intermediate network devices due to software bugs or delayed acknowledgments. \emph{Jitter} is a fluctuation in delay between source and sender nodes, which can be caused by routing anomalies or transient congestion on some parts of the network. We injected these anomalies at varying rates to show how the severity level affects the performance. For example, packet loss and corrupt are injected with $0.05\%$, $0.1\%$, $0.5\%$, and $1\%$ rates and reorder and duplicate are injected with $15\%$, $25\%$, $35\%$, and $45\%$ rates. We also reproduced network congestion using a separate client-server pair that share a common link with the original transfer task. 


As can be observed in Figure~\ref{fig:net_cong_a}, packet loss and delay anomalies severely degrade the performance of transfers. The impact of packet corruption and reorder is noticeable but not as severe as loss and corrupt anomalies. On the other hand, duplicate anomaly does not affect the throughput which could be attributed to network capacity being larger than I/O read/write performance. Similar to loss and corrupt anomalies, congestion also results in significant performance drops for the file transfer. Hence, different anomalies can result in the same outcome, thus it is not possible to predict the root cause simply based on throughput values. 

\textbf{File System:} HPC clusters utilize parallel file systems (PFS) to provide high-performance I/O. PFSes are typically shared by all compute and transfer nodes, creating a risk of resource contention when multiple users/jobs access the same storage node to read or write files at the same time \cite{yildiz2016root,li2015ascar}. For example, network contention can occur when data paths for two I/O operations share the same switch or storage device. Likewise, congestion can arise when there are more requests than a storage or client node can handle efficiently \cite{li2015ascar}. This, in turn, can result in I/O performance becoming a potential bottleneck for file transfers.
To illustrate the impact of storage node congestion on the performance of file transfers, we created two Lustre clusters each consisting of one metadata server, two object storage servers, and two client nodes. We then used one of the clients in each Lustre cluster to transfer $30\times3$GiB dataset between the clusters. While the transfer is running, we used the second client to create read/write congestion on the storage servers that the transfer application is using to read/write files. Figure~\ref{fig:read_write_cong} shows how the read/write I/O congestion on the source and destination clusters affects the throughput of the transfer. We increased the number of processes used to congest the storage node to simulate an increased level of congestion. It is clear that read/write I/O congestion on both ends of the transfer results in a significant throughput decrease for the transfer. 

\textbf{Data Transfer Node:} Misconfiguration and resource contention on DTNs adversely affect the performance of data transfers. Since most transfer applications (e.g., FTP, scp, and GridFTP) do not require high CPU and memory resources for file transfers, inducing memory and CPU congestion does not lead to considerable throughput degradation. On the other hand,  TCP buffer size misconfiguration can severely limit the performance since TCP requires buffer size to be  equal to or greater than bandwidth-delay product (BDP) to achieve full network utilization. Yet, it is not uncommon to find HPC clusters with misconfigured TCP buffer size settings. For example, maximum TCP buffer size is set to $6$, and $8$ MiB in Bridges-2~\cite{bridges} and Expanse~\cite{expanse} supercomputers, respectively, which is significantly smaller than necessary to reach full network utilization. Figure~\ref{fig:tcp_buff} demonstrates the impact of TCP send and receive buffer size values on transfer throughput on a network with $120$ KiB Bandwidth Delay Product (BDP), thus setting TCP buffer size to smaller values causes transfer throughput to decrease considerably. For example, setting receiver side buffer size to $2$KiB constraints the transfer throughput to only $40$Kbps. 

\section{End-to-end File Transfer Monitoring Framework}
\begin{figure}
\begin{center}
\includegraphics[width=1\linewidth]{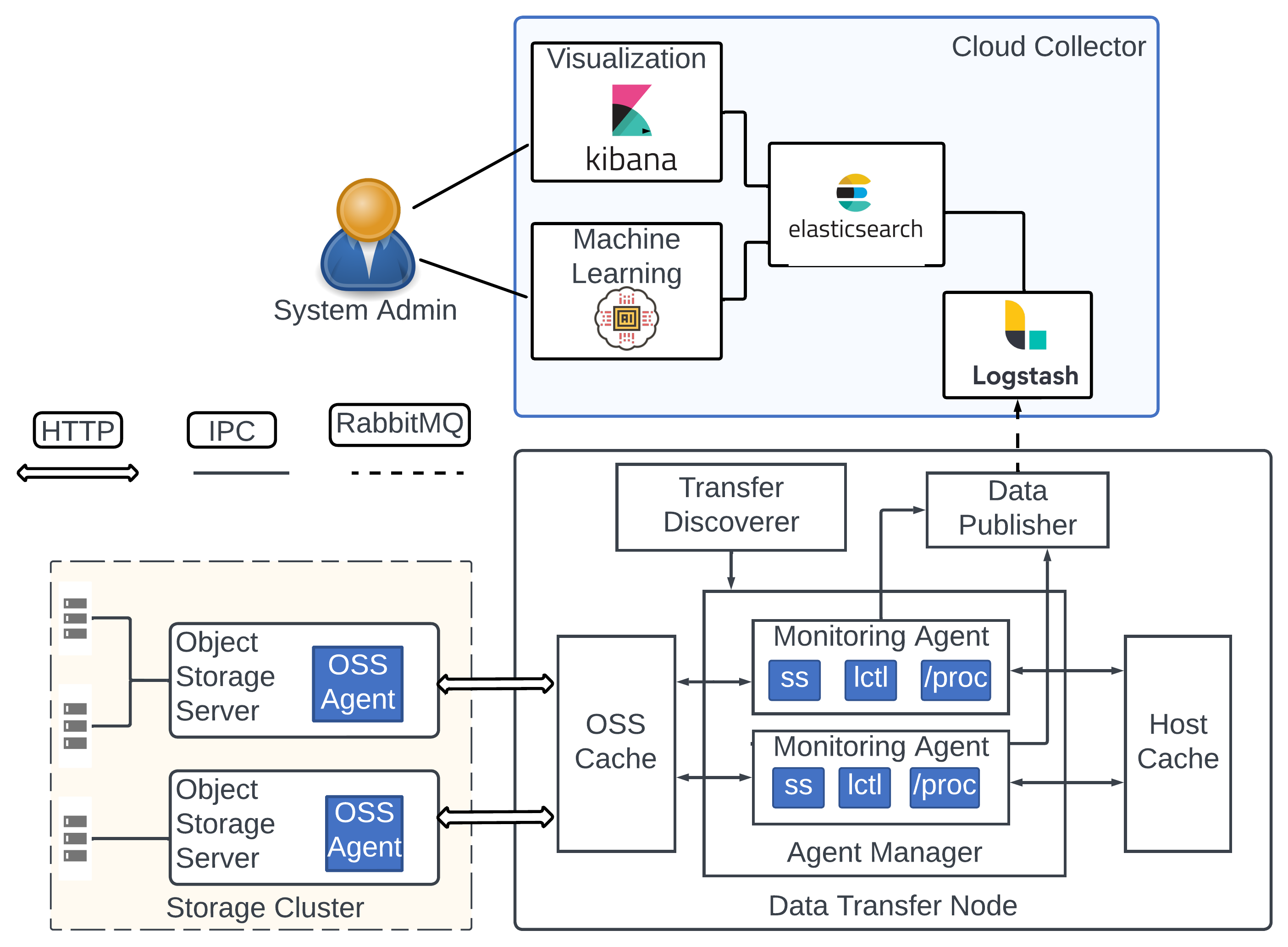}
\caption{Overview of the proposed monitoring framework. Monitoring agents running on data transfer nodes gather performance metrics for the file system, network, and DTN and push them to a central Data Collector for analysis, visualization, and storage.}
\label{fig:architecture}
\end{center}
\vspace{-6mm}
\end{figure}


A high-level architecture of the proposed monitoring framework is illustrated in Figure~\ref{fig:architecture}. It comprises seven services, of which five run on a data transfer node, one on a storage server, and the last at a cloud datacenter. \emph{Transfer Discoverer}, \emph{Agent Manager}, \emph{Monitoring Agent}, \emph{OSS Cache}, \emph{Host Cache}, and \emph{Data Publisher} are the services that operate on data transfer nodes. \emph{OSS Agent} runs on storage nodes and serve I/O statistics to \emph{OSS Cache} when requested. Lastly, \emph{Cloud Collector} runs in a cloud datacenter and processes messages sent by \emph{Data Publisher(s)}. 

\textit{Transfer Discoverer} periodically (by default once a second) scans active TCP connections (using Linux \texttt{ss} utility) to detect new transfers as well as to notice completed transfers. Once it finds that a new transfer has started, it messages to the \textit{Agent Manager} to spawn a Monitoring Agent to monitor the resource usage of the transfer. Please note that the Transfer Discoverer searches for both incoming and outgoing transfers to capture performance statistics on both ends of transfers. To avoid tracking short-lived web traffic flows, the Transfer Discoverer adopts a whitelisting approach to only keep track of transfers that run between predefined IP address ranges. If Transfer Discoverer cannot find a currently monitored transfer among active connections for two consecutive scanning periods, it signals to Agent Manager to terminate the Monitoring Agent assuming that the transfer is completed. 
\begin{table}
\begin{centering}
\caption{Sample performance metrics collected by monitoring agents to pinpoint performance anomalies.}
\label{tab:sample_metrics}
\renewcommand{\arraystretch}{0.9} 
\begin{tabular}{p{1.4cm} p{2.2cm} p{4cm}}
\hline
\textbf{Component} & \textbf{Metric Name}  & {\bf Description} \\
\hline
Storage &ost\_read\_bytes &  {Total amount of read size in OST}\\
\hline
Storage &ost\_write\_bytes &  {Total amount of write size in OST}\\
\hline
DTN & snd\_buffer\_max&  {TCP maximum send buffer size}\\
\hline
DTN & rcv\_buffer\_max& {TCP maximum receive buffer size}\\
\hline
DTN & NIC\_send\_bytes& {Send bytes of the DTN Lustre NIC}\\
\hline
DTN & NIC\_receive\_bytes & {Receive bytes of the DTN Lustre NIC}\\
\hline
Network & segs\_out&  {The number transmitted segments}\\
\hline
Network & segs\_in&  {The number received segments}\\
\hline
Network & average\_rtt &   {Average round trip time} \\
\hline
Network & retrans&  {The number of retransmitted packets}\\
\hline
\end{tabular}
\end{centering}
\vspace{-5mm}
\end{table}

\begin{table*}[!h]
\begin{centering}
\caption{Specification of testbeds used to reproduce performance anomalies.}
\vspace{-3mm}
\label{table:1}
\begin{tabular}{p{2cm}  p{3.5cm}  p{3cm}   p{2.5cm}  p{3.5cm}} 

\bf  &\bf Storage &\bf CPU &\bf  Memory &\bf Network\\
\hline
\bf Testbed \#1 &  
SATA SSD &
20 cores @ 2.60 GHz  &
160 GiB &
10 Gbps, 0.2 ms RTT \\
\hline
\bf Testbed \#2 & 
SATA HDD &
20 cores @ 2.60 GHz  &
160 GiB &
10 Gbps, 0.2 ms RTT \\
\hline
\bf Testbed \#3 & 
SATA HDD \& SATA SSD &
20 cores @ 2.60 GHz  &
160 GiB &
10 Gbps, 0.2 ms RTT \\
\hline
\bf Testbed \#4 & 
SATA SSD &
20 cores @ 2.60 GHz  &
160 GiB &
10 Gbps, 10 ms RTT\\
\hline
\bf Testbed \#5 & 
SATA HDD &
20 cores @ 2.60 GHz  &
160 GiB &
10 Gbps, 10 ms RTT\\
\hline
\bf Testbed \#6 & 
SATA SSD &
16 cores @ 3.00 GHz  &
128 GiB &
25 Gbps, 0.2 ms RTT\\
\hline
\bf Testbed \#7 & 
SATA SSD &
16 cores @ 3.00 GHz  &
128 GiB &
25 Gbps, 30 ms RTT\\
\hline
\bf Testbed \#8 & 
SATA SSD &
16 cores @ 3.00 GHz  &
128 GiB &
25 Gbps, 10 ms RTT\\
\hline
\end{tabular}
\end{centering}
\vspace{-4mm}
\end{table*}

\textit{Monitoring Agent} is responsible for collecting performance metrics for active transfers. The data collection period of this service is configurable with a default value of one second. Without the loss of generality, we designed the Monitoring Agent architecture to gather performance statistics from Lustre clusters since Lustre is the most commonly used file system in HPC clusters. Although Lustre has many components that can adversely affect the I/O performance, in this work, we only focus on detecting resource interference events that take place at the client or storage server sides. On the Lustre client side, interference can happen when multiple transfers or compute jobs (running on the same data transfer node) read/write files (either from the same OST or different OST) at the same time, congestion the link between the data transfer node and LNET router. On the Lustre storage server side, congestion take place when multiple transfers or compute jobs access the same Object Storage Target simultaneously. Hence, the Monitoring Agent collects performance metrics for the Lustre clients and OSSes. 

Please note that a transfer process reads/writes from/to a single file at an given time hence Monitoring Agent only communicates to one Object Storage Server at any given time. Thus, we identify the storage server that is used to read/write the transfer file and only capture statistics for that storage server instead of capturing entire file system statistics, which will unnecessarily induce significant overhead. To do so, we first identify the file that the transfer application is currently operating on by reading the \texttt{/proc/fd} file for the transfer process. Then, we execute the \texttt{lfs} command to find the OST where the file is located. Finally, we send a request to the OSS Cache service to access the performance counters for the identified OST by communicating to the OSS responsible for the identified OST. 

To access data transfer node (i.e., DTN) statistics, Monitoring Agents communicate with the Host Cache service which collects node-level metrics including TCP buffer size value and and Lustre Object Storage Client and Metadata Client statistics. Finally, we utilize \emph{ss} utility to capture several key network metrics for file transfers including round trip time, total and retransmitted packet counts. Once all the metrics are collected, the Monitoring Agent passes them to the Data Publisher. It utilizes (IPC) when communicating OSS Cache, Host Cache, and Data Publisher services. Monitoring Agent captures a total of $142$ metrics, out of which $55$ metrics are related to I/O performance, $16$ metrics are related to network performance, and $71$ metrics are related to DTN performance. Table~\ref{tab:sample_metrics} presents sample metrics collected for individual components.


\textit{OSS Cache} service communicates to OSS Agents running on Object Storage Servers to gather performance statistics for Object Storage Targets. \textit{Host Cache} service captures performance counters for the data transfer node using \texttt{psutil} and \texttt{lctl} utilities. While it is possible that Monitoring Agents capture OST and DTN metrics themselves, we observed that this incurs significant overhead as the number of transfers increases due to increase number of concurrent requests. \textit{OSS Agent} is a lightweight monitoring process that runs on Object Storage Servers to collect performance metrics for Object Storage Targets. It utilizes \texttt{lctl} command to access OST metrics periodically (once a second, by default) and make them available OSS Cache which serves them to Monitoring Agents when requested. It keeps the messages in a key-value cache such that Monitoring Agent(s) can access them quickly when needed. We define a time limit for the cache to remove old logs to minimize memory footprint.


\textit{Data Publisher} service receives performance metrics from all Monitoring Agents running on a node and pushes them to the pub/sub message queue. \textit{Cloud Collector} service receives the messages sent from the Data Publishers and saves them in the database using Elastic Search, LogStash, and Kibana (ELK) stack. LogStash pulls messages from RabbitMQ and inserts them into the Elastic Search database. Finally, Kibana pulls data from Elastic Search for visualization.

\begin{figure*}
\begin{center}
    \subfigure[Monitoring Agent Performance - 142 metrics]{
    \includegraphics[keepaspectratio=true,angle=0,width=.4\textwidth] {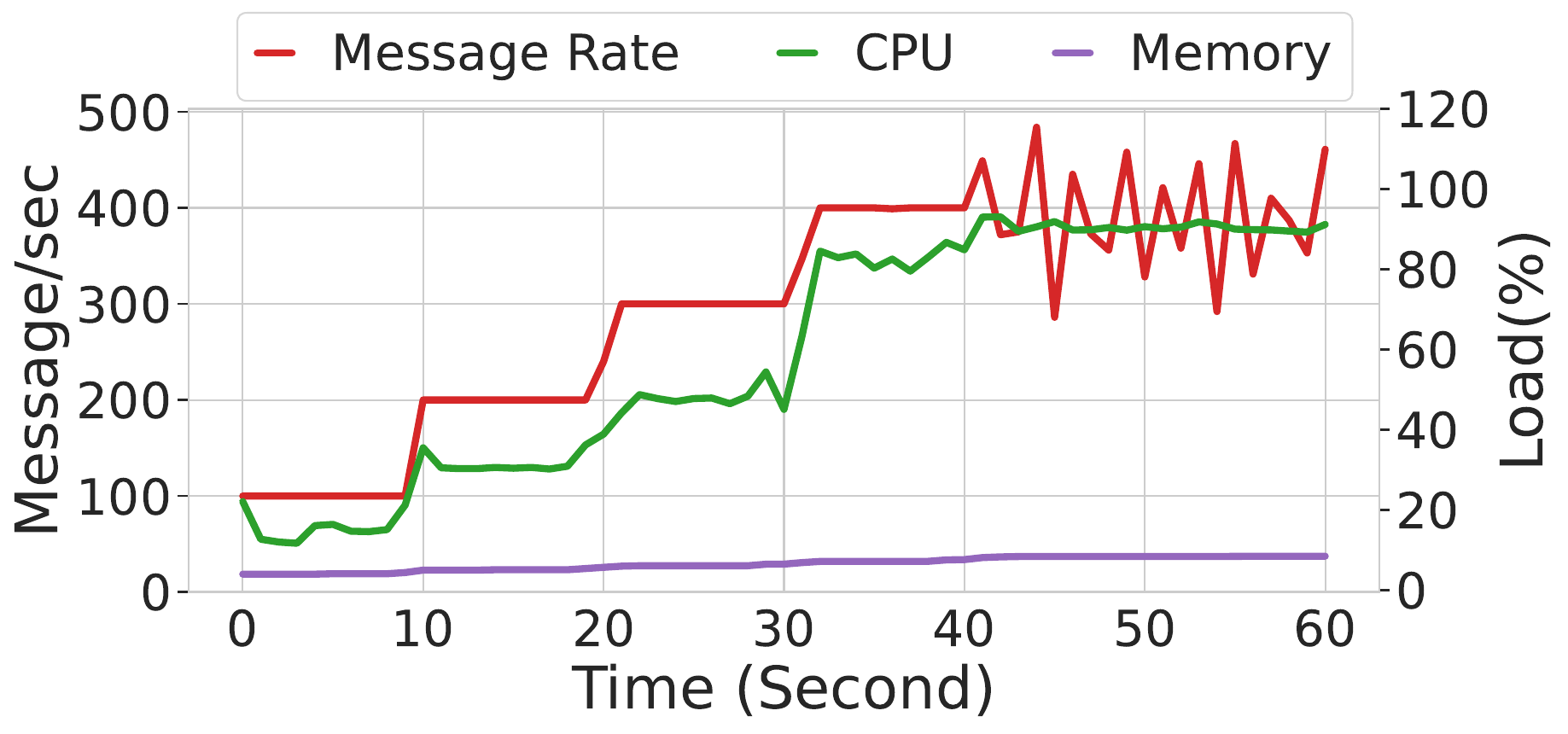}
    \label{fig:scalability_agent}}
    \hspace{3mm}
    \subfigure[Monitoring Agent Performance - 14 metrics]{
    \includegraphics[keepaspectratio=true,angle=0,width=.4\textwidth] {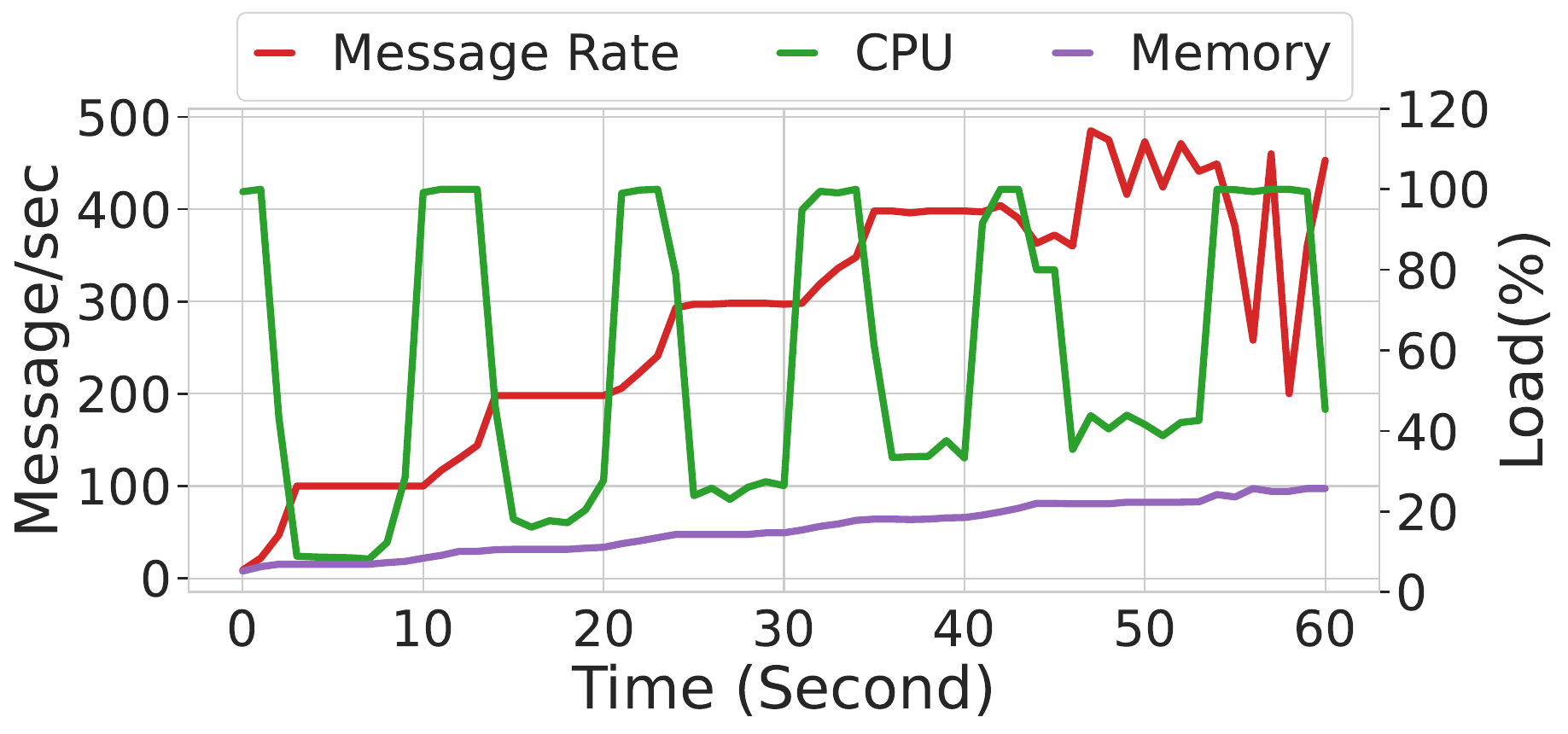}
    \label{fig:scalability_agent_14metrics}}
    \hspace{3mm}
    \subfigure[Cloud Collector Performance -142 metrics]{
    \includegraphics[keepaspectratio=true,angle=0,width=.4\textwidth] {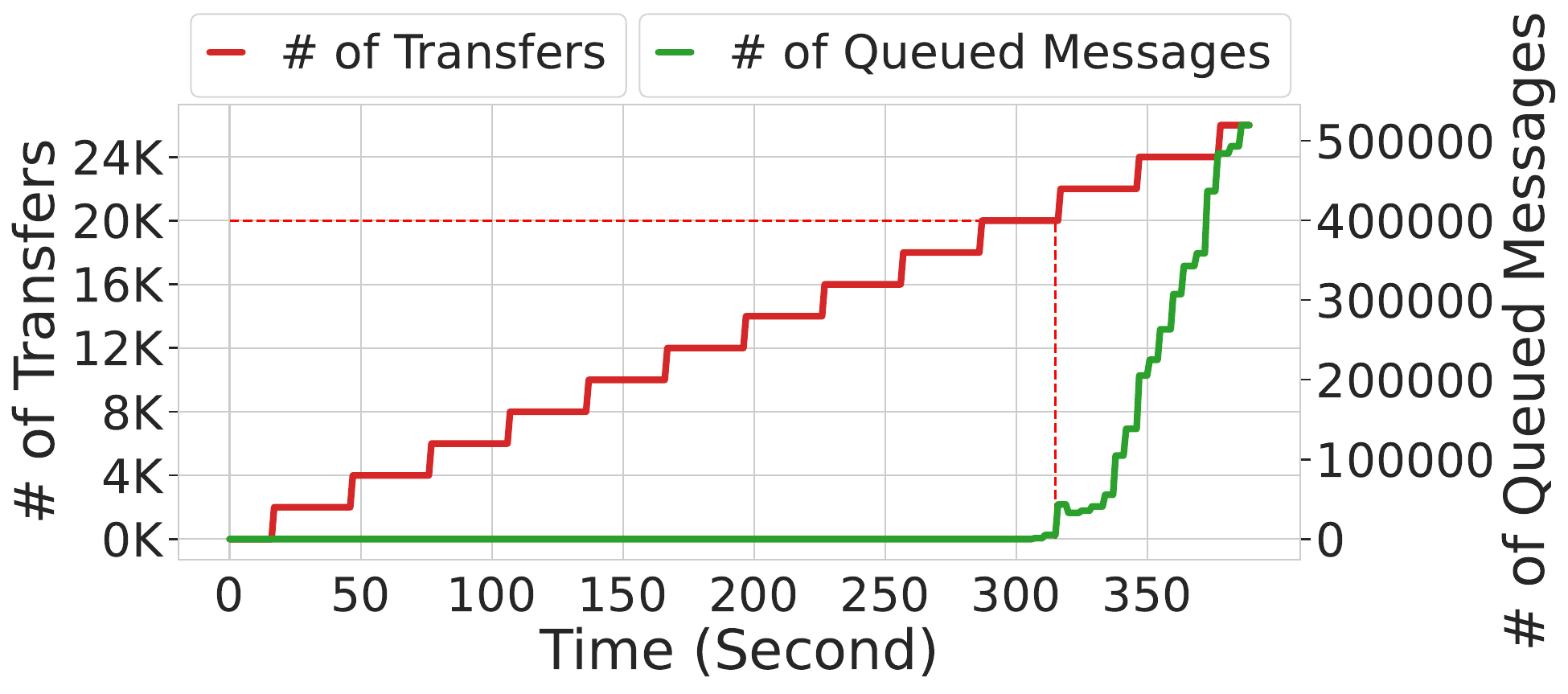}
    \label{fig:scalability_cloud}}
    \subfigure[Cloud Collector Performance - 14 metrics]{
    \includegraphics[keepaspectratio=true,angle=0,width=.4\textwidth] {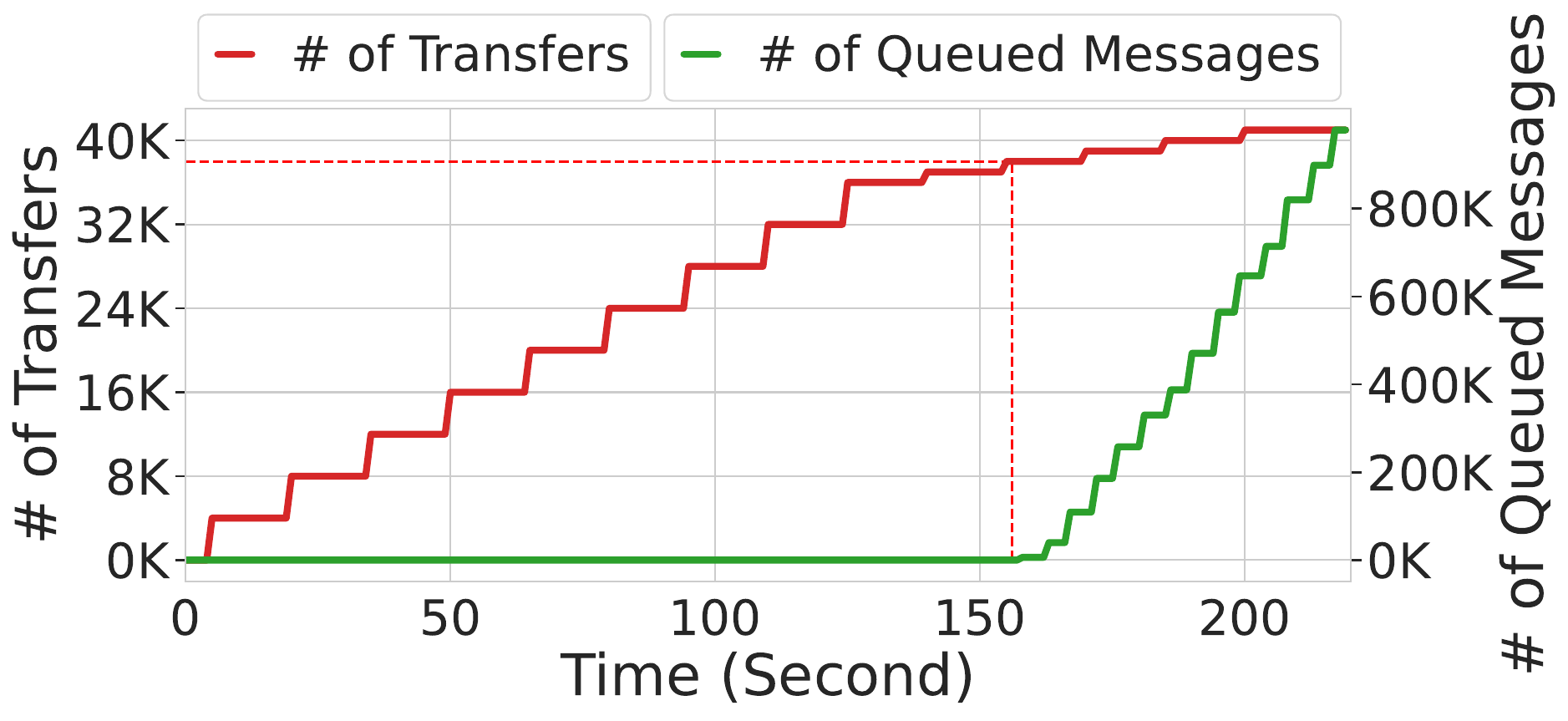}
    \label{fig:scalability_cloud_14metrics}}
\caption{Scalability analysis of the proposed monitoring framework. It can monitor up to $400$ transfers in one server and up to $40K$ transfers in total.}
\vspace{-6mm}
\label{fig:scalability}
\end{center}
\end{figure*}

To facilitate the visualization and analysis of collected data for each transfer, we assign a unique transfer ID created by combining time (i.e., year, month, day, hour, minute) and connection (i.e., source IP address, source and port number, destination IP address and destination port number) information. The transfer ID is then used as a key when pushing data to Elastic Search. This helps to combine performance statistics from source and destination end points of the same transfer task as they both sender and receiver nodes pick the same unique transfer ID. Since it is possible that sender and receiver nodes are located in different timezones, we use UTC time on both ends to ensure synchronization. It is also important to note that the proposed monitoring framework is transfer application agnostic. In other words, it does not require any modification in transfer applications as it discovers all necessary information by itself. For example, it utilizes a series of utilities to identify which file is being transferred and in which storage node it is stored at. This in turn results in slightly longer execution time and performance overhead in exchange for supporting more transfer applications. To validate this, we run and monitor GridFTP transfers for scalability analysis results (i.e., Figure~\ref{fig:scalability}).

\section{Performance Evaluation}


We evaluate the performance of the presented model in terms of scalability, usability, and prediction accuracy. Unless otherwise specified, all results are presented with a data collection frequency of one second. We created eight clusters each consisting of at least $14$ nodes, seven on the sender side and seven on the receiver side. Out of seven nodes, one is allocated for Lustre metadata and management services, three are used as OSSes, and the last three are used as client/data transfer nodes. Client nodes have at least two network interfaces; one is used by the transfer application to send/receive the data between sender/receiver nodes other one is used to stream data in/out of Lustre. Specifications of these clusters are given in Table~\ref{table:1}. The testbeds differ from each other in terms of disk types (SSD vs HDD), disk capacity, network bandwidth, and delay.




\subsection{Overhead Analysis}
We first assess the execution time and resource usage of a single Monitoring Agent in the proposed framework. We measure the execution duration, total data size sent to Data Publisher, CPU usage, and memory usage. We calculate these values by running the proposed framework on a data transfer node in Testbed-1 to collect performance metrics when monitoring a single transfer in one second intervals. We captured these metrics on both sender and receiver nodes and report the largest values in Table~\ref{tab:overhead_measurement}. We measured the metrics for two variations of the monitoring agent implementation. One version collects $142$ performance metrics from all the components involved (i.e., storage server, storage client, metadata client, transfer node, and network connection) for both sender and receiver ends. The other version only collects $14$ metrics that are sufficient to identify performance anomalies, as discussed in Section~\ref{sec:prediction}.

\begin{table}
\renewcommand{\arraystretch}{1.2}
\begin{centering}
\caption{Execution time and resource usage of the proposed framework when monitoring single transfer application.}
\label{tab:overhead_measurement}
\resizebox{.45\textwidth}{!}{
\begin{tabular}{p{3cm} p{2.5cm} p{2.5cm}}
\hline
{\bf Metric} & {\bf 142 metrics} & {\bf 14 metrics} \\
\hline
Execution Time & 76ms &  14ms \\
\hline
Payload Size &  838B & 160B \\
\hline
CPU Utilization & $2\%$ & $0.2\%$ \\
\hline
Memory Usage & 20MiB & { 20 MiB} \\
\hline
\end{tabular}}
\end{centering}
\vspace{-6mm}
\end{table}

\begin{figure}
\begin{center}
\subfigure[142 metrics]{
\includegraphics[keepaspectratio=true,angle=0,width=.35\textwidth] {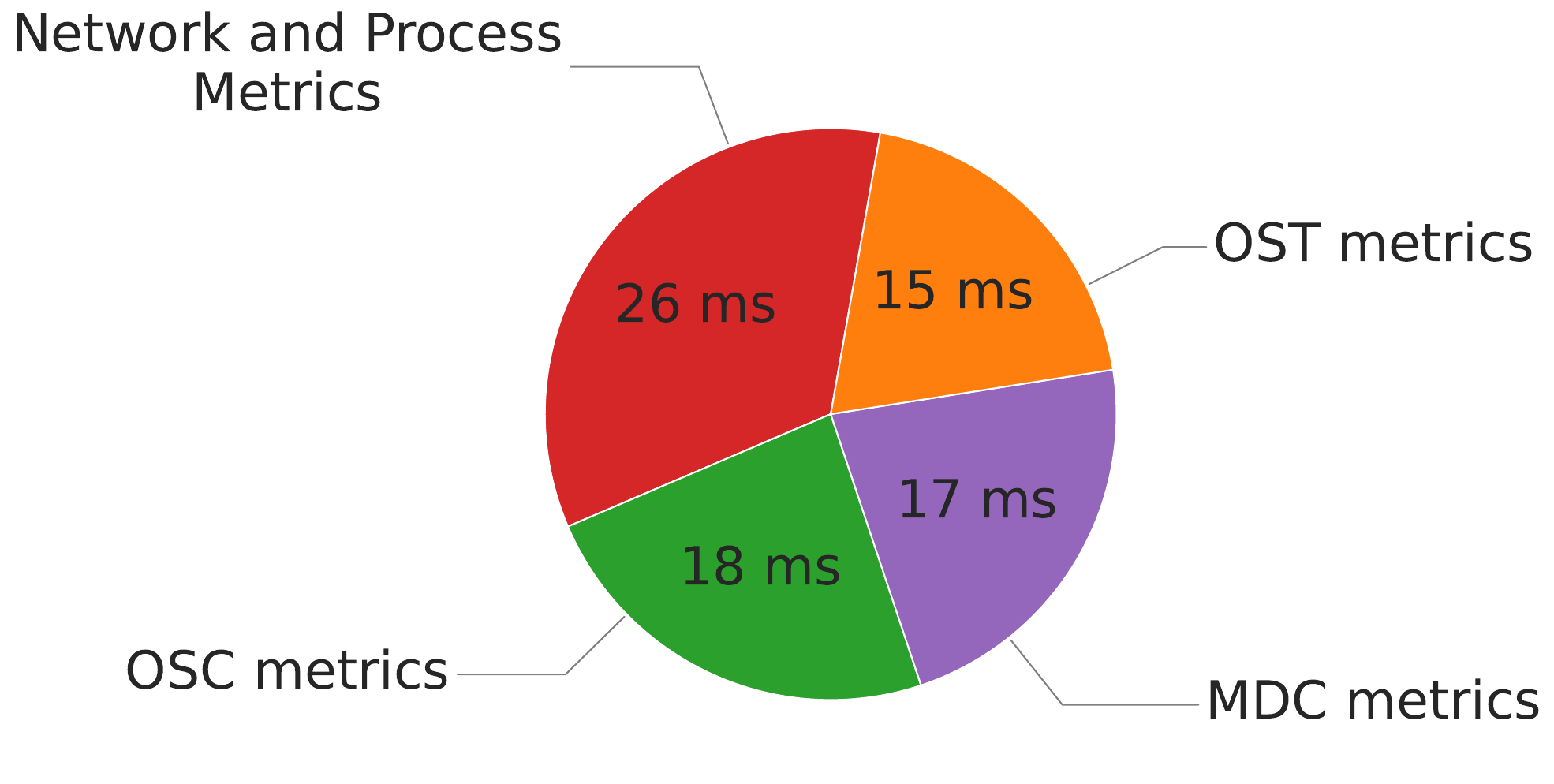}
\label{fig:execution_time_pie_chart}}
\subfigure[14 metrics]{
\includegraphics[keepaspectratio=true,angle=0,width=.35\textwidth] {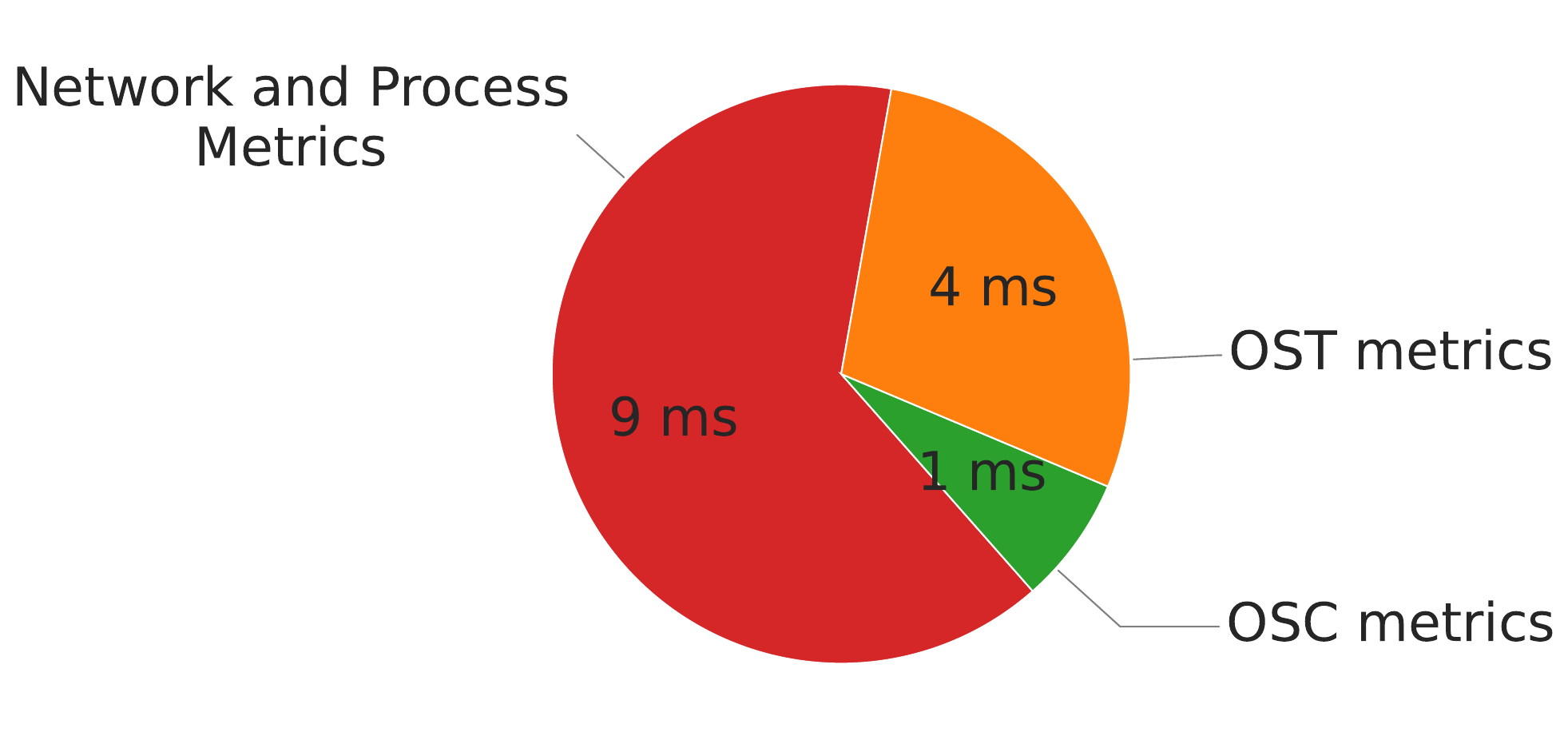}
\label{fig:execution_time_pie_chart_14metrics}}
 \caption{Execution time details for Monitoring Agents. It takes $76$ms to capture all performance counters and $14$ms to capture $14$ selected metrics.}   
\end{center}
\vspace{-8mm}
\end{figure}

As it can be observed, both the memory and CPU footprint of the Monitoring Agent is reasonably low with $20$ MiB memory usage and $0.2-2\%$ CPU usage on a single core depending on how many metrics are collected. To measure network usage, we also measured the message payload that is sent to the Data Publisher every second. We find that the message size is between $160-830$ bytes, which we believe is negligible compared to the network capacity of HPC clusters. Its execution time to gather performance metrics in each data collection interval is around $14ms$ when collecting the subset of metrics and $76$ms when it is collecting all $142$ metrics. The breakdown of execution time is given in Figure~\ref{fig:execution_time_pie_chart} and ~\ref{fig:execution_time_pie_chart_14metrics}. Monitoring Agent gathers both network and transfer process statistics by executing command line commands, which takes around $26ms$ when collecting all metrics. It then uses \texttt{lfs} utility to find the Lustre client and OST indexes that the transfer file is stored at. It then contacts Host Cache and OSS Cache services to obtain OSC, MDC, and OST metrics. It takes $15-18$ ms to run and capture these metrics as well. Cache services (OSS Cache and Host Cache) play a significant role in keeping the data collection time low mainly because they prefetch the data and keep them in memory such that Monitoring Agents can access the data quickly. We observed that the runtime for Monitoring Agents reaches over $400$ms when cache servers are removed since it requires Monitoring Agents to communicate to OSS Agents located on storage servers and execute commands in real time.

\begin{figure*}
\begin{center}
    \hspace{-3mm}
    \subfigure[Normal Transfer]    {\includegraphics[keepaspectratio=true,angle=0,width=.32\linewidth]{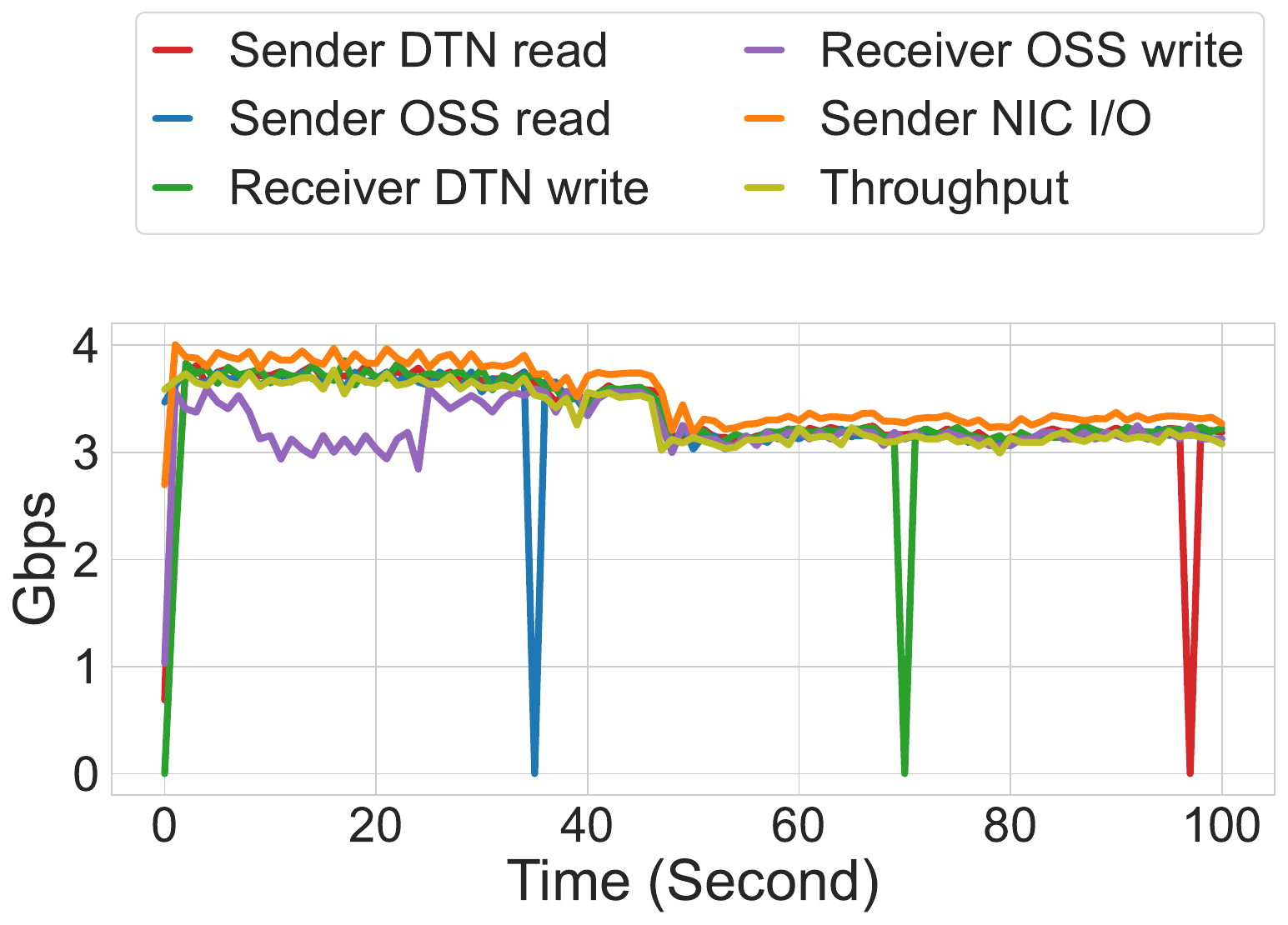}
    \label{fig:no_cong_a}}
    \hspace{-3mm}
    \subfigure[Receiver Lustre Client Congestion]    {\includegraphics[keepaspectratio=true,angle=0,width=.32\linewidth]{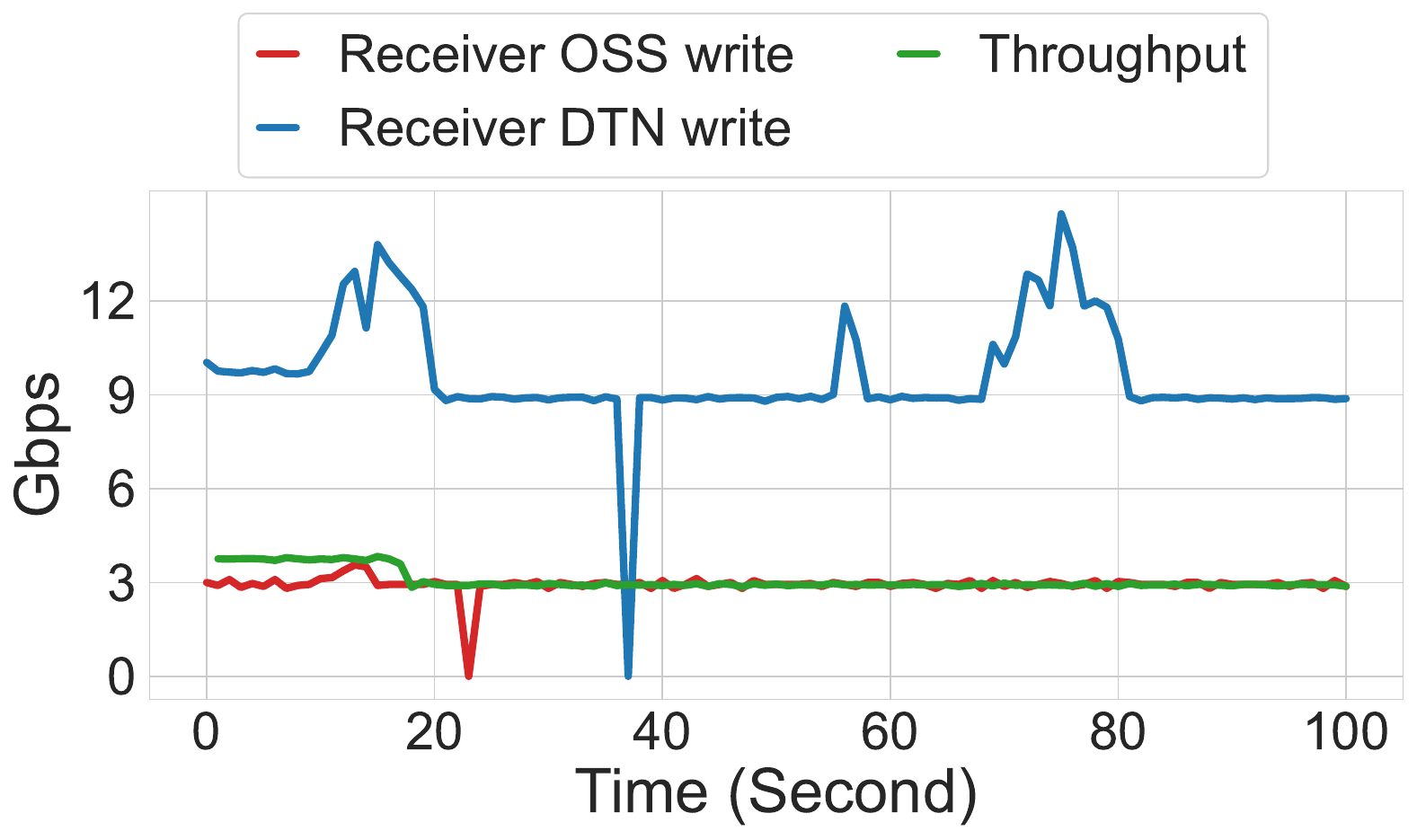}
    \label{fig:write_cong_b}}
    \hspace{-3mm}
     \subfigure[Receiver Lustre Server Congestion]    {\includegraphics[keepaspectratio=true,angle=0,width=.32\linewidth]{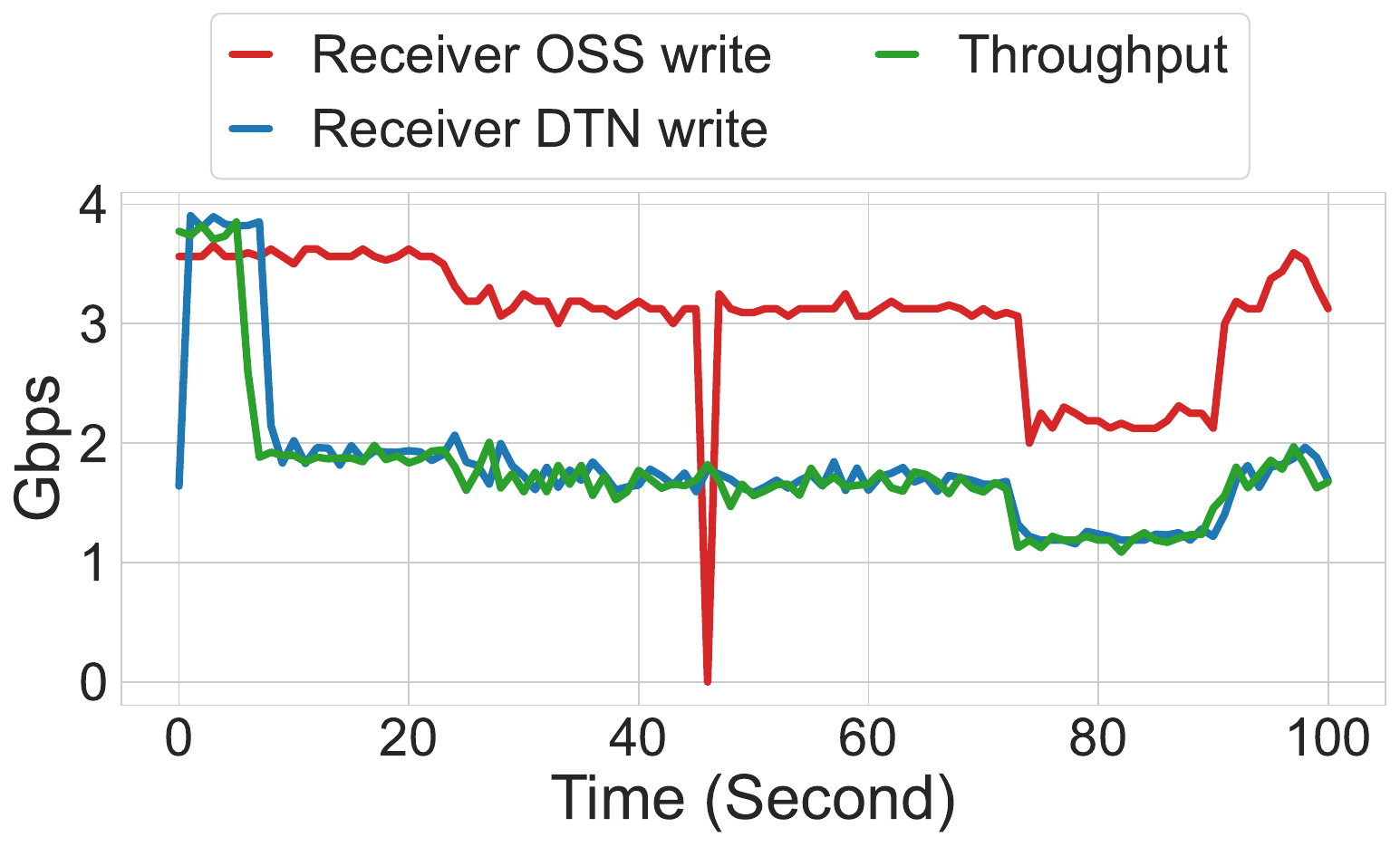}
    \label{fig:write_cong_c}}
    \hspace{-3mm}   
    
     \subfigure[Sender Lustre Client Congestion]    {\includegraphics[keepaspectratio=true,angle=0,width=.32\linewidth]{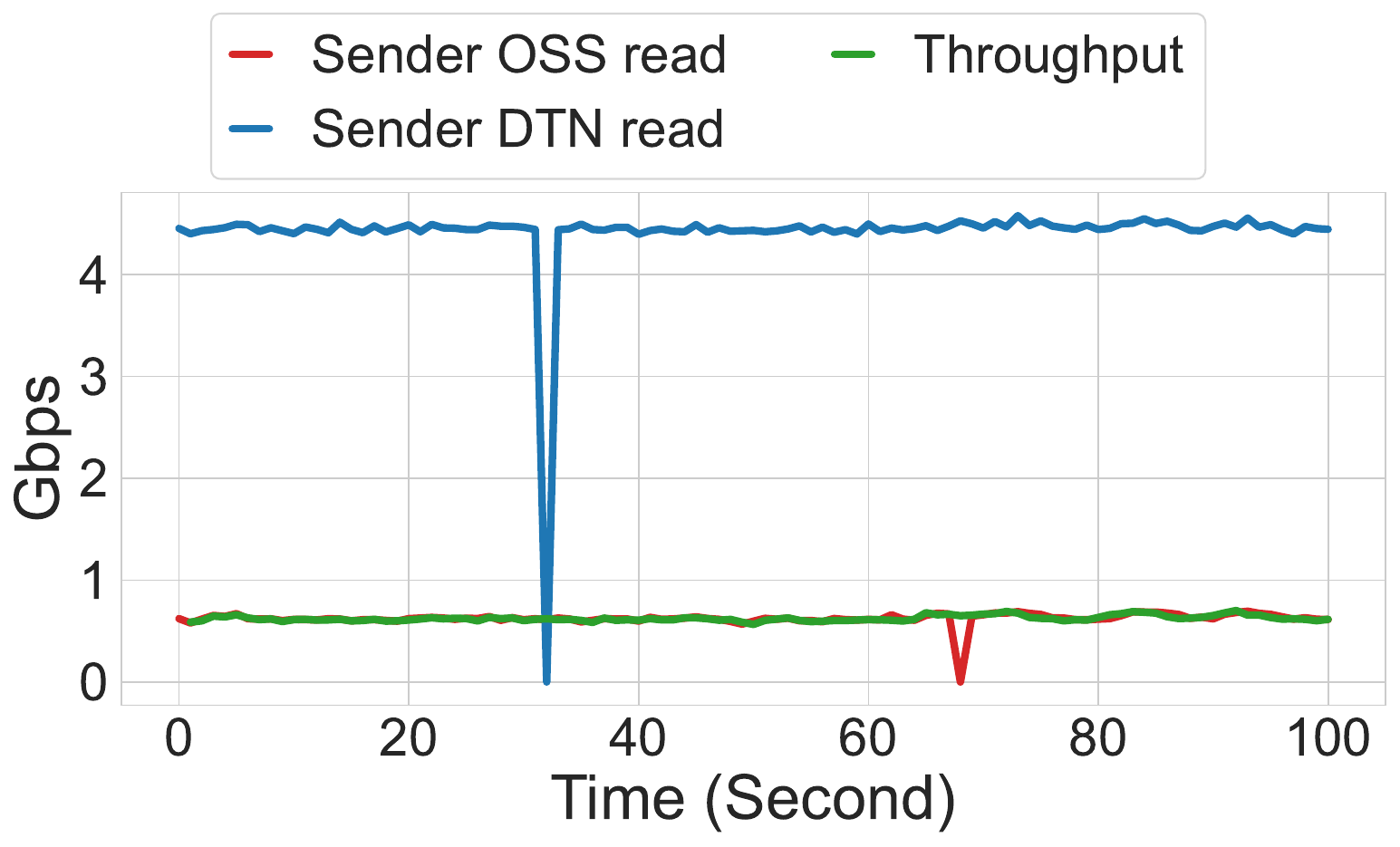}
    \label{fig:read_cong_d}}
    \hspace{-3mm} 
     \subfigure[Sender Lustre OSS Congestion]    {\includegraphics[keepaspectratio=true,angle=0,width=.32\linewidth]{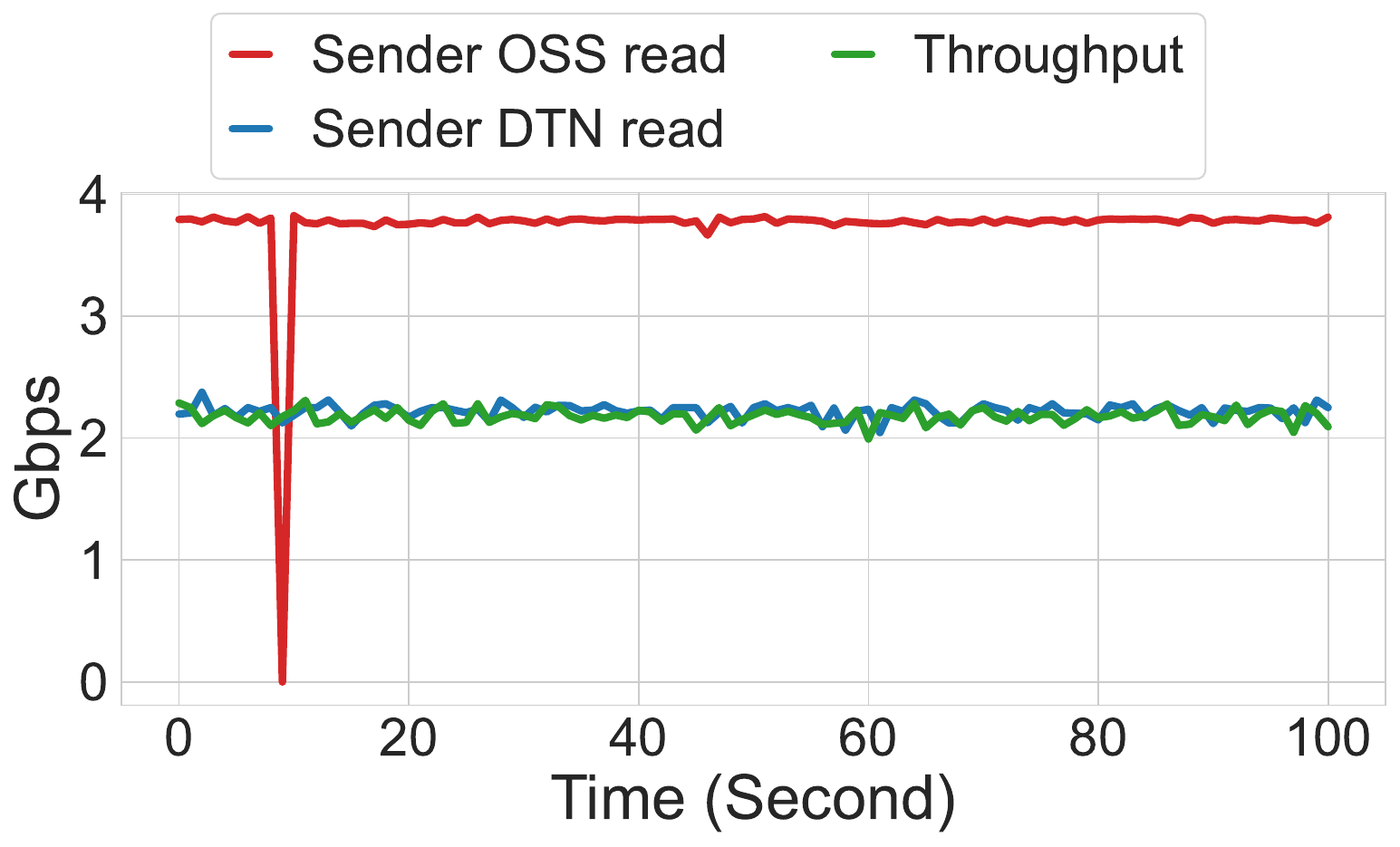}
    \label{fig:read_cong_e}}
    \hspace{-3mm}  
     \subfigure[Network Congestion]    {\includegraphics[keepaspectratio=true,angle=0,width=.32\linewidth]{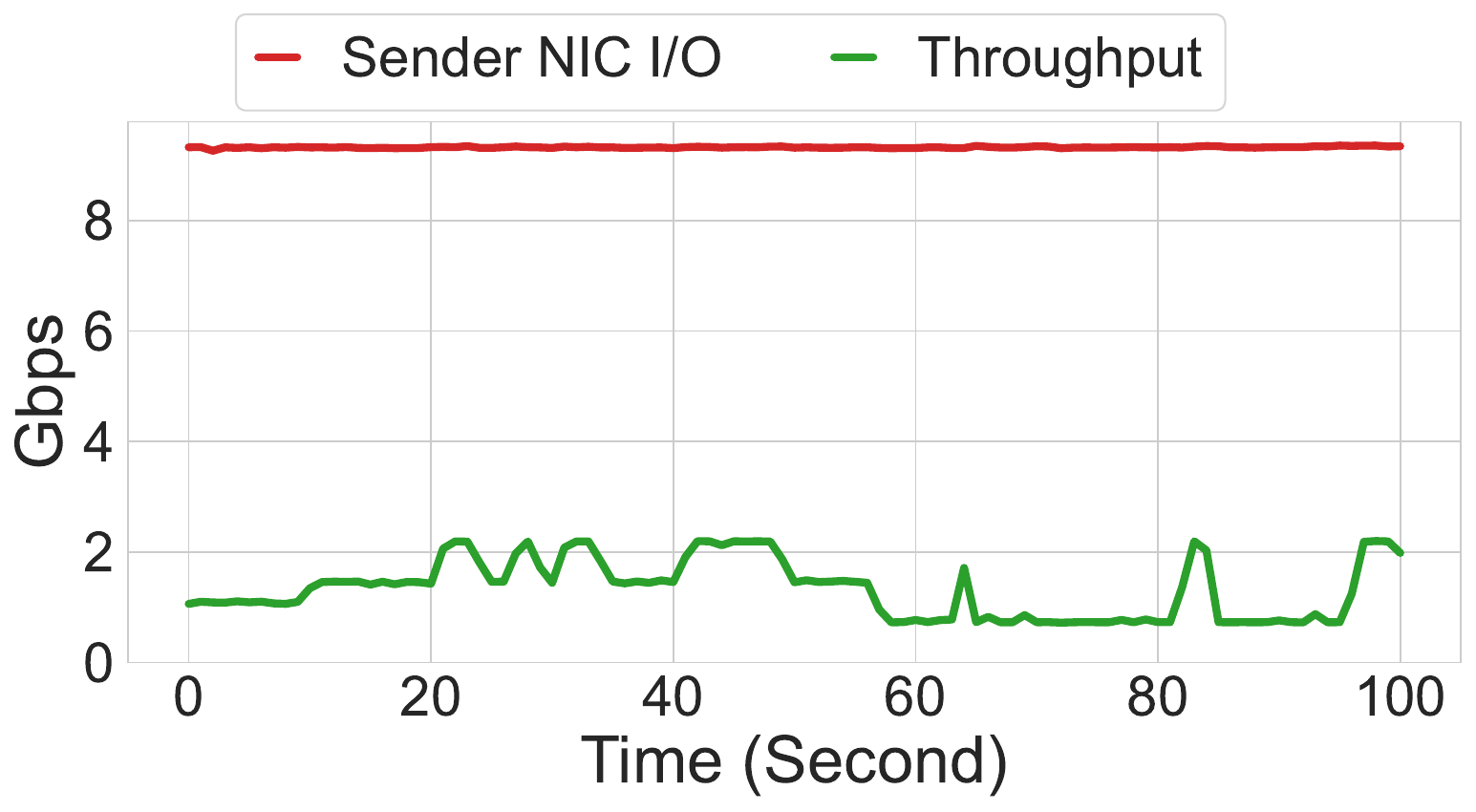}
    \label{fig:net_cong_f}}
    \hspace{-3mm}   
    \label{fig:anomaly_visuals}
\caption{Illustration of various performance metrics under normal and abnormal scenarios. The proposed solution can be used to visualize the collected performance metrics in real-time to facilitate performance debugging.}
\vspace{-5mm}
\end{center}
\end{figure*}
\vspace{-2mm}
\subsection{Scalability Analysis}
 Figure~\ref{fig:scalability} presents the scalability results for the Monitoring Agent and Cloud Collector services. In Figure~\ref{fig:scalability_agent} and~\ref{fig:scalability_agent_14metrics}, we initiate $100$ transfers every $10$ seconds and monitor the number of messages published by Data Publisher along with CPU and memory utilization of the data transfer node. We observe that the number of messages increases at a similar rate with an increasing number of transfers up to $400$ transfers.
 This indicates that Monitoring Agents can gather performance statistics within the data collection interval (i.e., one second) and send them to Cloud Collector to be processed without causing performance bottlenecks. Average CPU utilization on all cores is around $80\%$ when $400$ transfers are monitored and $142$ metrics are sent to Cloud. This value decreases to around $40\%$ when collecting $14$ metrics. Similarly, the memory footprint is around $5-20\%$ which corresponds to around $5-20GiB$ as the total memory size on the node is $128$ GiB. Please note that CPU and memory usage values are reported for the whole system, including the cached files that are sent on the sender side. Thus actual usage of Monitoring Agents is less than these values. Specifically, while Monitoring Agents are likely responsible for most of the CPU usage, they account for less than $50\%$ of memory usage.

To evaluate the performance of Cloud Collector, we deployed the ELK stack on a c6a.16xlarge instance on Amazon Web Service cloud provider, which has $64$ AMD EPYC cores, $128$ GiB memory, and $25$ Gbps network interface card speed. To emulate the monitoring of a large number of active transfers from multiple data transfer nodes, we use $9$ data transfer nodes, each publishing performance metrics for an increasing number of transfers from $200$ to $4,000$. Since it is not feasible to run this many transfers on a single data transfer node, Data Publisher simply pushes the same data for a defined number of transfers. We observe that the ELK stack on Cloud Collector can receive and store up to $20,000$ messages per second without any performance issues when $142$ performance metrics are monitored and sent to the Cloud. When the number of transfers reaches more than $20,000$, we observe build-up in the message queue mainly because of the performance limitations of the Elastic Search database.
If the monitoring agents only send $14$ performance metrics to the cloud, the ELK stack on the Cloud Collector can handle close to $40,000$ transfers per second without any performance problems. If we assume that an average data transfer node handles $20$ transfers at any given time, a single Cloud Collector instance will be able to serve $1,000-2,000$ data transfer nodes. Please note that it is also possible to increase the capacity of Cloud Collector even more through horizontal and vertical scaling.


\subsection{Performance Debugging} \label{sec:visualization}

\newcolumntype{P}[1]{>{\centering\arraybackslash}p{#1}}
\begin{table*}
\begin{centering}
\caption{F1 score of machine learning models when predicting the root causes of performance anomalies using 12 features. The Random Forest model attains close-to-maximum performance with the lowest training cost.}
\label{table:models_perf_a_b_c}
\begin{tabular}{p{2.6cm}  P{1.5cm}   P{1.5cm} P{1.5cm}   P{1.5cm} P{1.5cm}   P{1.5cm} P{1.5cm}   P{1.5cm} }
\hline
\bf ML Model & \bf Testbed \#1 & \bf Testbed \#2 & \bf Testbed \#3 & \bf Testbed \#4 &\bf Testbed \#5 &\bf Testbed \#6 &\bf Testbed \#7 &\bf Testbed \#8\\ 
\Xhline{1pt}

\bf Random Forest & 100.0 & 99.69 & 99.7 & 99.67 & 99.72 & 99.86 & 99.57 & 99.91\\
\bf Decision Tree &  99.84 & 99.52 & 99.29 & 99.38 & 99.3 & 99.74 & 99.39 & 99.7\\
\bf XGBoost & 99.95 & 99.6 & 99.5 & 99.72 & 99.78 & 99.98 & 99.77 & 99.84 \\
\bf Neural Network & 99.93 & 99.6 & 99.58 & 99.6 & 99.65 & 99.86 & 99.34 & 99.67 \\
\bf Support Vector M. & 99.76 & 99.18 & 99.41 & 99.41 & 99.18 & 99.88 & 98.99 & 99.49 \\
\hline
\end{tabular}
\end{centering}
\end{table*}
In this section, we discuss how the proposed monitoring framework can allow system administrators to conveniently debug performance anomalies. For this analysis, we transferred the $20\times3GiB$ dataset using Testbed \#1. To eliminate the effect of Lustre client and server-side caching, we clear the cache before we start a transfer.
Figure \ref{fig:no_cong_a} shows multiple I/O and network metrics for a normal transfer scenario. While the average throughput is around $3.3$ Gpbs, it drops to $2.9$ Gbps when there is write congestion on the data transfer node (Figure \ref{fig:write_cong_b}). By comparing ``Receiver DTN write'' (which indicates the throughput of a link between the DTN and LNet router), and ``Receiver OSS write'' metrics, one can infer that OSS is not under heavy load but the link between the DTN and LNet router is fully utilized. Receiver-side OSS write congestion happens when the OST that the transfer application is using is overloaded, hence the transfer application attains suboptimal I/O throughput. This can be understood from Figure~\ref{fig:write_cong_c} since Receiver OSS write values are close to the disk's maximum throughput (i.e., 3.4 Gbps) and it is larger than the amount of data written from the transfer node (i.e., Receiver DTN write). Hence, other applications must be issuing I/O operations to the same storage node, causing OSS congestion. Similar observations can be made for sender-side I/O congestion issues (Figure \ref{fig:read_cong_d} and \ref{fig:read_cong_e}) to identify the root cause of performance issues.

Finally, we investigate the network congestion scenario which can take place more often compared to other network anomalies like packet corruption and reordering. For this case, we first congest the link between the sender and the receiver by running multiple \texttt{iPerf} \cite{iperf3} transfers alongside the target transfer. As can be seen in Figure \ref{fig:net_cong_f}, the sender NIC throughput reaches close-to-maximum ($10$ Gbps) while the transfer throughput is less than $2$Gbps throughput. 



\subsection{Automated Bottleneck Detection} \label{sec:prediction}

While it is possible to manually debug some transfer issues, it is not feasible to do the same for all transfers due to the lack of human power and the relatively time-consuming process of comparing multiple metrics to identify the root cause. Hence, an automated solution that can process the real-time data and predict the bottleneck of transfers is needed to take full advantage of monitoring and mitigate performance issues in a timely manner.

\begin{figure}
\centering
    \subfigure[Without Feature Normalization]{\label{fig:tr_before}
        \includegraphics[width=42mm]{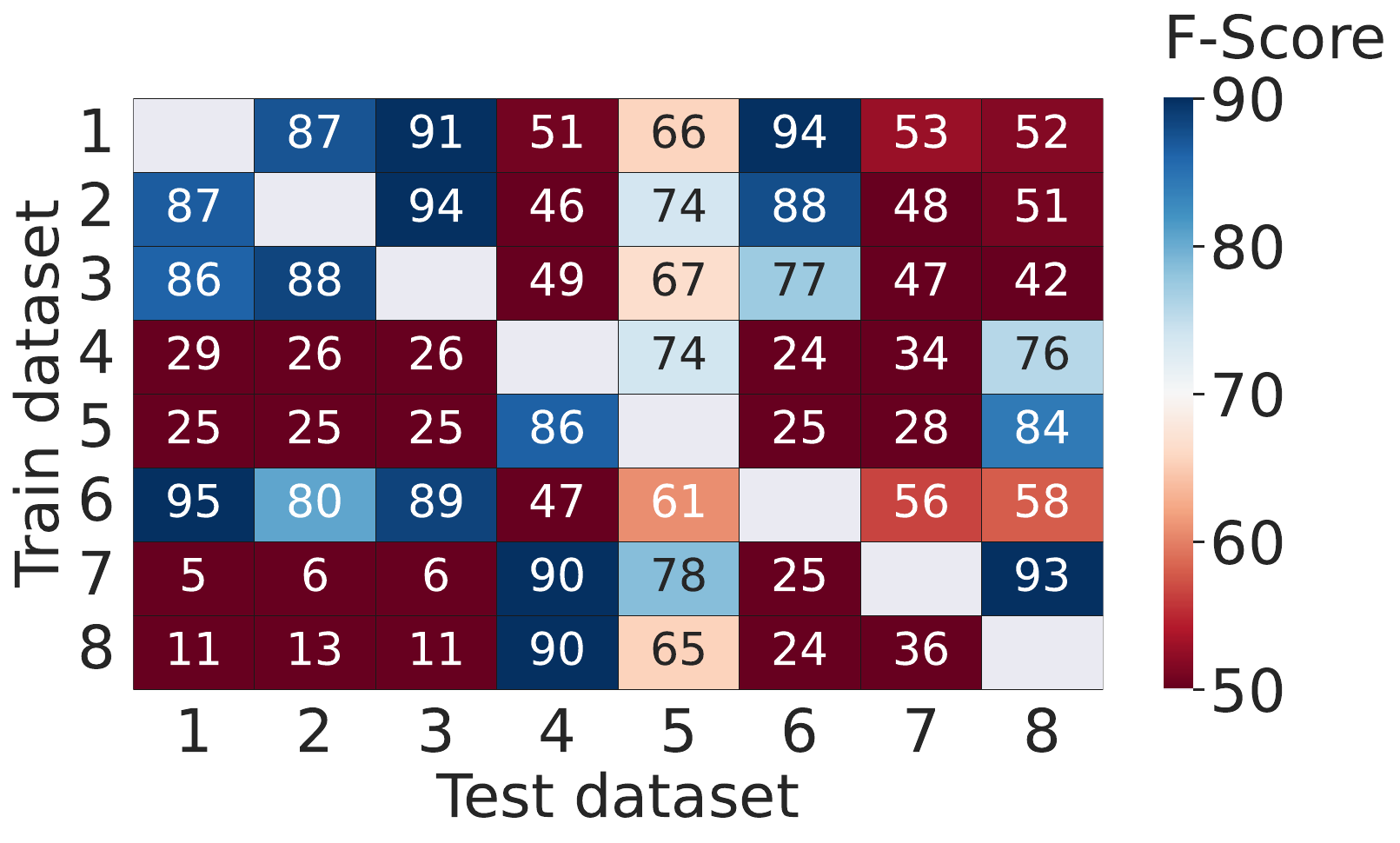}
    }
    \hspace{-4mm}
    \subfigure[With Feature Normalization]{\label{fig:tr_after}
        \includegraphics[width=42mm]{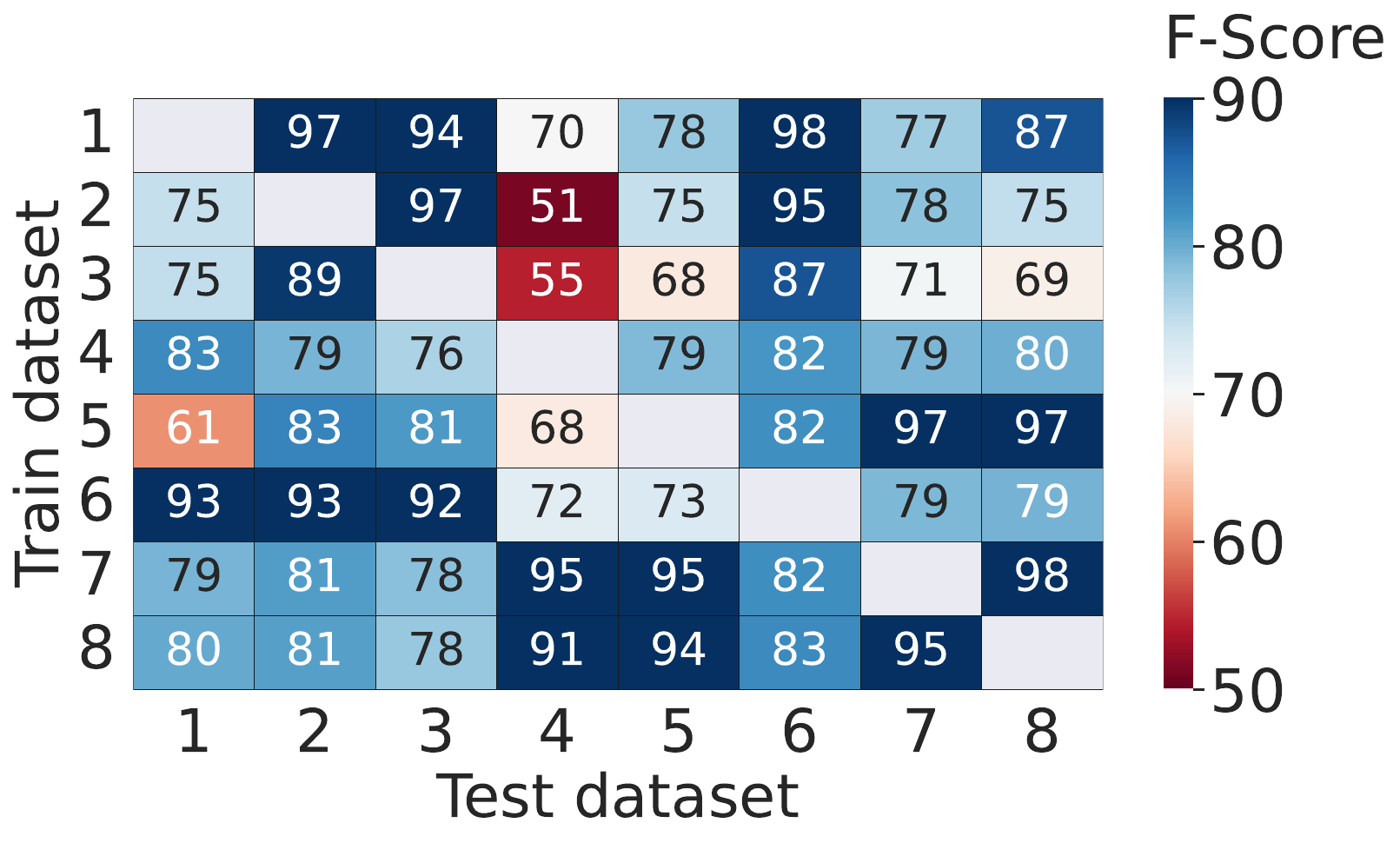}
    }
\vspace{-3mm}
\caption{Transfer learning performance of Random Forest model  with and without feature normalization. 
}
\vspace{-6mm}
\end{figure}

Since it is not possible to reproduce transfer anomalies in production systems, we created our own datasets to develop an automatic bottleneck prediction solution. On the sender side, read I/O interference is one of the most likely problems that can limit the transfer throughput. It can happen at the OST level when multiple Lustre clients access to same OST. The interference can also occur at the DTN level when multiple transfer applications simultaneously access files located in the same or different OSTs such that even though individual OSTs are not congested, the Lustre client network interface on the DTN may be overloaded. As a result, we consider two different types of I/O congestion cases as OST read and DTN read. Since I/O congestion can also happen on the receiver side, we also reproduced OST write and DTN write congestion scenarios on the receiver end. We also reproduced the TCP buffer size misconfiguration scenario, as it is a common mistake in many production systems. To do so, we set the maximum TCP buffer size to smaller than bandwidth-delay product values both on the sender and receiver ends. Finally, we considered two network anomalies as loss and congestion. As a result, $4$ I/O contention, $2$ DTN misconfiguration, and $2$ network anomaly/congestion anomalies are reproduced.



To simulate I/O congestion, we run separate processes to read/write data from/to the file system while transfers are running. We modify the maximum TCP buffer size setting before starting transfers for TCP buffer size anomalies. Finally, we use Linux \texttt{netem}~\cite{tcnetem} utility to manually configure the Linux kernel packet scheduler to emulate network anomalies. To avoid the impact of caching during transfers, we clear the cache memory by forcing the operating system to drop all cache data on all engaged nodes in a transfer before starting a transfer. As illustrated in Figure~\ref{fig:anomaly_examples}, we reproduced anomalies in different severity levels. We chose the severity levels such that the throughput of the transfers drop by the $20-80\%$ range such that we can cover a wide range of the possible outcomes of congestion issues. 

\newcolumntype{P}[1]{>{\centering\arraybackslash}p{#1}}

\begin{table*}
 \renewcommand{\arraystretch}{1.4}
\begin{centering}
\caption{Heuristic approach to process performance metrics to determine the root cause of low transfer throughput.}
\label{table:4}
\begin{tabular}{p{3.2cm}  p{6cm}   p{7.5cm} } 

\hline
\bf Bottleneck Type & \bf Conditions  & \bf Justification \\ 
\Xhline{1pt}
    Receiver side OST write congestion & 
    1. OST write bytes $\approx$ Average of OST write bytes in normal case\newline 2. Transfer receiver write bytes $<$ Receiver OST write bytes &
    This happens when OST is utilized close to maximum capacity (condition 1) but the transfer receiver is not  the sole client that writes to target OST (condition 2)\\
\hline
    Receiver side Lustre client congestion &
    1. Receiver write bytes  $<$  Receiver write bytes in normal class\newline 2. Receiver's Lustre NIC interface sent bytes $>$ Receiver's Lustre NIC interface sent bytes in the normal class & 
    This happens when the write size of the Lustre client on the receiver DTN is smaller than its range in the normal case and the Receiver's Lustre NIC interface sends more data than its normal range.\\
\hline
    Sender side OST read congestion&
    1. Sender OST read bytes $\approx$ Average of sender OST read bytes in normal case\newline 2. Sender lustre client read bytes $<$ Sender OST read bytes &  
    This happens when the read bytes of the Lustre client on the sender DTN is less than the read bytes of the OST on the OSS server, and the read bytes metric of the OST on the OSS server is within its maximum read range.\\
\hline
    Sender side Lustre client congestion &
    1. Sender lustre client read bytes  $<$  Sender lustre client write bytes in normal class\newline 2. Sender's Lustre NIC interface received bytes$>$ Sender's Lustre NIC interface received bytes in normal class&  
    This happens when the read bytes of the Lustre client on the sender DTN  is smaller than the normal label, and Sender's Lustre NIC interface receives more data than the normal.\\
\hline
    Sender TCP Buffer Value Misconfiguration &
    1. TCP send buffer value on the sender side $\ll$ Bandwidth Delay Product &  
    This happens when TCP maximum buffer size on the sender node is set to smaller than BDP.\\
\hline
    Receiver TCP Buffer Value Misconfiguration &
    1. TCP receive buffer value on the receiver side $\ll$ Bandwidth Delay Product & 
    This happens when TCP maximum buffer size on the receiver node is set to smaller than BDP. \\
\hline
    Network Loss &
    1. Ratio of retransmitted packets to total sent packets is more than $0.05$\% & 
    This happens when a transfer experiences more than usual packet loss rate \\
\hline
    Network Congestion &
    1. Round trip time $>$ 1.5 * Round trip time in normal class \newline 2. Receiver Lustre client write bytes $<$ Receiver Lustre client write bytes in normal class & 
    In a congested network packets would be delivered in a larger time than in a non-congested network. Thus, we expect to observe an increase in the round trip time due to queue build-up. We also expect the throughput of the transfer to decrease considerably.\\
\hline
\end{tabular}

\end{centering}
\end{table*}

\subsubsection{Machine Learning Models}
We first trained multiple machine learning models to predict the eight anomaly/congestion cases. To evaluate the performance of models, We measured F-1 scores, which is a harmonic mean of precision and recall values~\cite{sokolova2009systematic}. Since our data is low-dimensional, we applied lightweight ML models that are faster to train for a real-time system and better fit for classification tasks~\cite{taheri2023deep}. We trained Support Vector Machine (SVM), Neural Network, Decision Tree, Random Forest, and eXtreme Gradient Boosting Tree (XGBoost) models by splitting the dataset from each network as training ($80$\%) and testing ($20$\%). We train each model three times on each network and report the average F-1 score to measure its performance. As shown in Table~\ref{table:models_perf_a_b_c}, all models attain very high scores (more than $99\%$ F-score) even with default settings; thus, we did not execute an extensive hyperparameter tuning.

Next, we investigate the transferability of machine learning models, which defines the success of ML models when training and test datasets come from different networks. As an example, a model is trained using a dataset collected in Testbed \#1 and tested against the data collected in Testbed \#2. Transferability is quite important for the adoption of ML models since gathering training datasets in all networks (especially in production systems) is not feasible. Figure~\ref{fig:tr_before} shows the F1-score of the Random Forest models when they are trained in one testbed and tested against other testbeds. On average, models attain only $54.7\%$ F-score mainly because of the high domain dependence of ML models. Specifically, some of the collected metrics are dependent on hardware characteristics; thus, ML models fail to perform well when tested against datasets collected on other networks with different hardware settings. To overcome this challenge, we applied feature transformation to normalize transfer metrics based on the values in a normal class. Hence, when testing a model in a different network, we first normalize the collected metrics using values of normal class in the test network. To do so, we divide the values of each class by their corresponding values in the normal class. It helps us to avoid using raw values and measure the rate of changes in performance compared to normal transfers. Figure~\ref{fig:tr_after} shows the performance of the Random Forest models after applying the proposed feature normalization approach. We observe that the performance of the model is improved to $81.82\%$, on average. Please note that it requires one normal sample from the test network to normalize the test dataset, which we believe is a reasonable expectation. However, F-score is still very low (around $50\%$) in some cases even after feature transformation.

\newcolumntype{P}[1]{>{\centering\arraybackslash}p{#1}}
\begin{table*}
 \renewcommand{\arraystretch}{1.2}
\begin{centering}
\caption{Precision, Recall, and F1-score of proposed heuristic-based classification approach when predicting the root causes of nine classes of performance anomalies. While ML models can attain $82\%$ F-score when they are transferred across networks, the heuristic model obtains more than $87\%$ in all networks.}
\label{table:heu_models_perf}
\begin{tabular}{p{2.3cm}  P{1.5cm}   P{1.5cm} P{1.5cm}   P{1.5cm} P{1.5cm}   P{1.5cm} P{1.5cm}   P{1.5cm} } 
\hline
\bf Metric & \bf Testbed \#1 & \bf Testbed \#2 & \bf Testbed \#3 & \bf Testbed \#4 &\bf Testbed \#5 &\bf Testbed \#6 &\bf Testbed \#7 &\bf Testbed \#8\\ 
\Xhline{1pt}

Precision & 98.20 & 94.36 & 94.12 & 89.82 & 91.07 & 97.37 & 94.13 & 95.44\\

Recall  & 98.08 & 92.50 & 92.35 & 86.53 & 88.37 & 97.10 & 92.39 & 90.19\\

F1-score  & 98.10 & 96.93 & 92.78 & 87.21 & 89.33 & 97.16 & 92.87 & 91.86\\

\hline
\end{tabular}

\end{centering}
\end{table*}

\subsubsection{Heuristic Approach}
To overcome the performance limitations of ML models and design an explainable solution, we introduced a heuristic bottleneck detection method. In this approach, we check the conditions that need to satisfy for a congestion (or anomaly) case to take place. For example, the rate of packet retransmission is expected to be high if the packet loss anomaly is experienced. Hence, we examined each anomaly and came up with a set of simple conditions to identify the presence of anomalies. Similar to the discussion in Section~\ref{sec:visualization}, storage node congestion can be deemed as a limiting factor for transfers if storage nodes are fully utilized, while the transfer client is only attaining a portion of this utilization. Table~\ref{table:4} lists the conditions along with reasons to mark a transfer with a certain category of anomaly. 
Table~\ref{table:heu_models_perf} presents the performance of the heuristic solutions. It attains $87-98$\% F-score in all networks, outperforming ML models significantly. Consequently, the heuristic model, despite being more labor-intensive to develop, offers a feasible option to process real-time performance statistics to make bottleneck predictions with high accuracy.

\section{Conclusion}
Despite significant investments to build high-performance research networks, the data transfers in these networks often fail to achieve high network utilization. Yet, there exists no solution to comprehensively monitor file transfers in research networks to understand the root causes of performance problems. In this paper, we make the first attempt to design a monitoring framework for file transfers that utilize lightweight monitoring agents to capture key performance statistics on data transfer nodes. The captured statistics are then transmitted to the cloud-hosted data collector for offline analysis as well as to process them in real-time with the help of a heuristic classification approach. Experimental results show that the proposed monitoring framework is highly scalable as it can monitor $400$ concurrent transfers per node and more than $40,000$ transfers in total. We also show that the proposed heuristic approach achieves up to $98\%$ F-1 score when evaluated with different levels of precision for root cause analysis. 

\vspace{-2mm}
\section*{Acknowledgement}
\vspace{-2mm}
The work in this study was supported in part by the NSF grants 2007789 and 2145742.


\bibliographystyle{IEEEtran}
\bibliography{main}

\end{document}